\documentclass[fleqn,usenatbib]{mnras}

\usepackage{newtxtext,newtxmath}
\usepackage[T1]{fontenc}

\DeclareRobustCommand{\VAN}[3]{#2}
\let\VANthebibliography\thebibliography
\def\thebibliography{\DeclareRobustCommand{\VAN}[3]{##3}\VANthebibliography}

\usepackage{graphicx}	% Including figure files
\usepackage{amsmath}	% Advanced maths commands
\usepackage{longtable}
\usepackage{caption}

\usepackage{color, colortbl}

\definecolor{LightCyan}{rgb}{0.88,1,1}
\definecolor{LightRed}{rgb}{1,1,0.88}

\newcommand{\micro}{\text{$\mu$}}

\newcommand{\jansky}{\text{Jy}}
\newcommand{\ujy}{\micro\text{Jy}}

\newcommand{\ergs}{\text{erg s$^{-1}$}}

\newcommand{\fluxden}{$\mu$\text{Jy\,beam$^{-1}$ }}

\newcommand{\bmins}{\ensuremath{.\hspace{-1.2mm}^{\prime}}}

\title[Terzan 5 Radio Imaging]{New Deep Radio Continuum Imaging Still Indicates a Large Reservoir of Undiscovered Millisecond Pulsars in Terzan 5}

\author[R. Urquhart et al.]{
Ryan Urquhart,$^{1}$\thanks{E-mail: ryan.t.urquhart@gmail.com}
Jay Strader,$^{1}$
Laura Chomiuk,$^{1}$
Scott M. Ransom,$^{2,3}$
Craig O. Heinke,$^{4}$
\newauthor
Arash Bahramian,$^{5}$
Thomas J. Maccarone$^{6}$
\\
$^{1}$Center for Data Intensive and Time Domain Astronomy, Department of Physics and Astronomy, Michigan State University, East Lansing, MI 48824, USA\\
$^{2}$National Radio Astronomy Observatory, 520 Edgemont Rd., Charlottesville, VA 22903, USA\\
$^{3}$Department of Astronomy, University of Virginia, Charlottesville, VA 22904, USA\\
$^{4}$Department of Physics, University of Alberta, CCIS 4-183, Edmonton, AB, T6G 2E1, Canada\\
$^{5}$International center for Radio Astronomy Research---Curtin University, GPO Box U1987, Perth, WA 6845, Australia\\
$^{6}$Department of Physics and Astronomy, Texas Tech University, Lubbock, TX 79409, USA\\}

\date{Accepted XXX. Received YYY; in original form ZZZ}

\pubyear{\the\year{}}

\begin{document}
\label{firstpage}
\pagerange{\pageref{firstpage}--\pageref{lastpage}}
\maketitle

\begin{abstract}
We present the deepest and highest-resolution radio continuum imaging of the Galactic globular cluster Terzan 5, one of the most crowded locations in the radio sky. In these new 2--4 GHz Karl G. Jansky Very Large Array images, we detect 38 of the 49 confirmed pulsars, including extensive multi-frequency eclipse mapping of the luminous redback Ter5A. Nonetheless, there is still a large amount of diffuse residual flux from pulsars that are fainter than our 2.5 GHz continuum detection limit of $\sim 11\,\mu$Jy. Using a range of approaches including image-based simulations, we model the fluxes of the detected pulsars together with the residual flux. We find a minimum total population of $N\sim250$ detectable pulsars in Terzan 5 and perhaps substantially more, though the luminosity function remains very uncertain. Consideration of the $\gamma$-ray properties of the cluster, though also not unambiguous to interpret, leads to consistent conclusions. These pulsar population estimates are larger than inferred from previous work and highlight Terzan 5 as a keystone target for next-generation radio facilities.
\end{abstract}

%%%%%%%%%%%%%%%%%%%%%%%%%%%%%%%%%%%%%%%%%%%%%%%%%%

%%%%%%%%%%%%%%%%% BODY OF PAPER %%%%%%%%%%%%%%%%%%

\section{Introduction} \label{sec:intro}

An early discovery from X-ray astronomy was a factor of $\sim 100$ overabundance of luminous X-ray sources in Galactic globular clusters compared to the field \citep{Katz1975}, including the idea that these are likely dynamically formed \citep{Clark1975}. Subsequent work established that these X-ray sources are mostly accreting neutron stars in binaries (e.g., \citealt{Grindlay1984}). The ensuing discovery of millisecond pulsars in globular clusters \citep{Lyne1987,Lyne1988} showed, at least in broad strokes, the expected progeny of dynamically formed X-ray binaries.

It was quickly realised that the origin of the neutron stars themselves is not necessarily straightforward. Neutron stars that form in core collapse supernovae receive natal kicks that are typically much higher than the escape velocities from cluster cores, so only a small fraction are expected to be retained---potentially too few to explain the observed population of millisecond pulsars (e.g., \citealt{Kulkarni1990,Pfahl2002}). This has led to models where the neutron stars in clusters are mainly formed by other mechanisms that produce lower kicks, such as electron capture supernovae 
(e.g., \citealt{Podsiadlowski2004}) or accretion-induced collapse of a white dwarf in a binary (e.g., \citealt{Ruiter2019}). Observations of X-ray binaries containing Be stars show evidence for a subpopulation of neutron stars that do indeed receive low ($\lesssim 10$ km s$^{-1}$) kicks due to a yet-unknown mechanism \citep{Pfahl2002b,Valli2025}. The physical processes that produce low-kick neutron stars in globular clusters likely depend on metallicity and other cluster parameters in a testable manner (e.g., \citealt{Ivanova2008}), though observational evidence for specific scenarios is still scant. 

The open questions do not stop at the origin of the old neutron stars recycled to form millisecond pulsars. \citet{Kirsten2022} reported the discovery of a fast radio burst in a metal-poor globular cluster in the nearby galaxy M81, which strongly suggests the presence of a very young, highly-magnetised neutron star in this old globular cluster. The close distance to M81 and the normalcy of this cluster (e.g., it has no unusual structural properties; \citealt{Dage2023}) implies that young neutron stars are forming in globular clusters at a meaningful rate, possibly through mergers of dynamically formed white dwarf--white dwarf binaries \citep{Kremer2021,Kremer2023}. Another recently discovered fast radio burst is localised with a large projected offset from a massive elliptical, most consistent with an old globular cluster host \citep{Shah2025}. These recent discoveries accompany existing evidence for a few young neutron stars in Galactic globular clusters (e.g., \citealt{Boyles2011,Kremer2024}). Overall, the ongoing uncertainty about the origin and fate of neutron stars in clusters motivates their continued study.

Terzan 5 has been a touchstone object in the study of neutron stars in globular clusters, hosting an early cluster X-ray burster \citep{Makishima1981} and found to contain only the second discovered eclipsing millisecond pulsar \citep{Lyne1990}. Radio continuum imaging with the Very Large Array (VLA) was used to infer the presence of many yet-undiscovered millisecond pulsars \citep{1990ApJ...365L..63F,2000ApJ...536..865F}, with the latter paper suggesting that Terzan 5 contained more pulsars than any other cluster, even though at that time only a single pulsar was known. This prediction was borne out by the discovery of 21 millisecond pulsars in Terzan 5 using the Green Bank Telescope by \citet{Ransom2005}. There are now 49 total pulsars confirmed in Terzan 5 \citep{Padmanabh2024}, indeed the largest known population of any cluster, including the fastest-spinning millisecond pulsar known in either cluster or field \citep{Hessels2006}. As all but one pulsar in Terzan 5 is a millisecond pulsar, we use the terms interchangeably in this paper.

The abundant pulsar population in Terzan 5 is a consequence of its high mass and dense core, which leads to an extraordinarily high stellar interaction rate---the highest of any Galactic globular cluster \citep{Bahramian2013}. However, the total population of pulsars in Terzan 5 is still very uncertain. Pulsar search observations can have difficulty detecting the eclipsing ``spider'' pulsars that are common in Terzan 5 as well as faint pulsars in close binaries, and are sensitivity limited even for isolated pulsars \citep{Cadelano2018,Padmanabh2024}.

In a previous paper (\citealt{2020ApJ...904..147U}, hereafter U20) we used new S (2--4 GHz) and C (4--8 GHz) band imaging of Terzan 5 obtained as part of the MAVERIC (Milky way ATCA and VLA Exploration of Radio sources in Clusters) survey of globular clusters \citep{Shishkovsky2020} to discover three new candidate spider millisecond pulsars. One of these has been subsequently confirmed as the redback pulsar Ter5ar \citep{Padmanabh2024}. However, neither the frequency band of these VLA data (split between S and C band) nor the spatial resolution (the data were taken in the hybrid BnA configuration, which is most similar to B configuration but with improved resolution for lower declination sources, like Terzan\,5) were optimised for studying pulsars.

Here we remedy these shortcomings with new, very deep S-band-only imaging of Terzan 5 in the VLA's highest-resolution A configuration, reaching rms noise levels and angular resolution both more than a factor of two better than existing data. These data allow us to both discover new candidate pulsars and set the best constraints yet on the total pulsar population of Terzan 5, indicating that hundreds of pulsars potentially await discovery.

\begin{figure}
    \centering
    \includegraphics[width=0.495\textwidth]{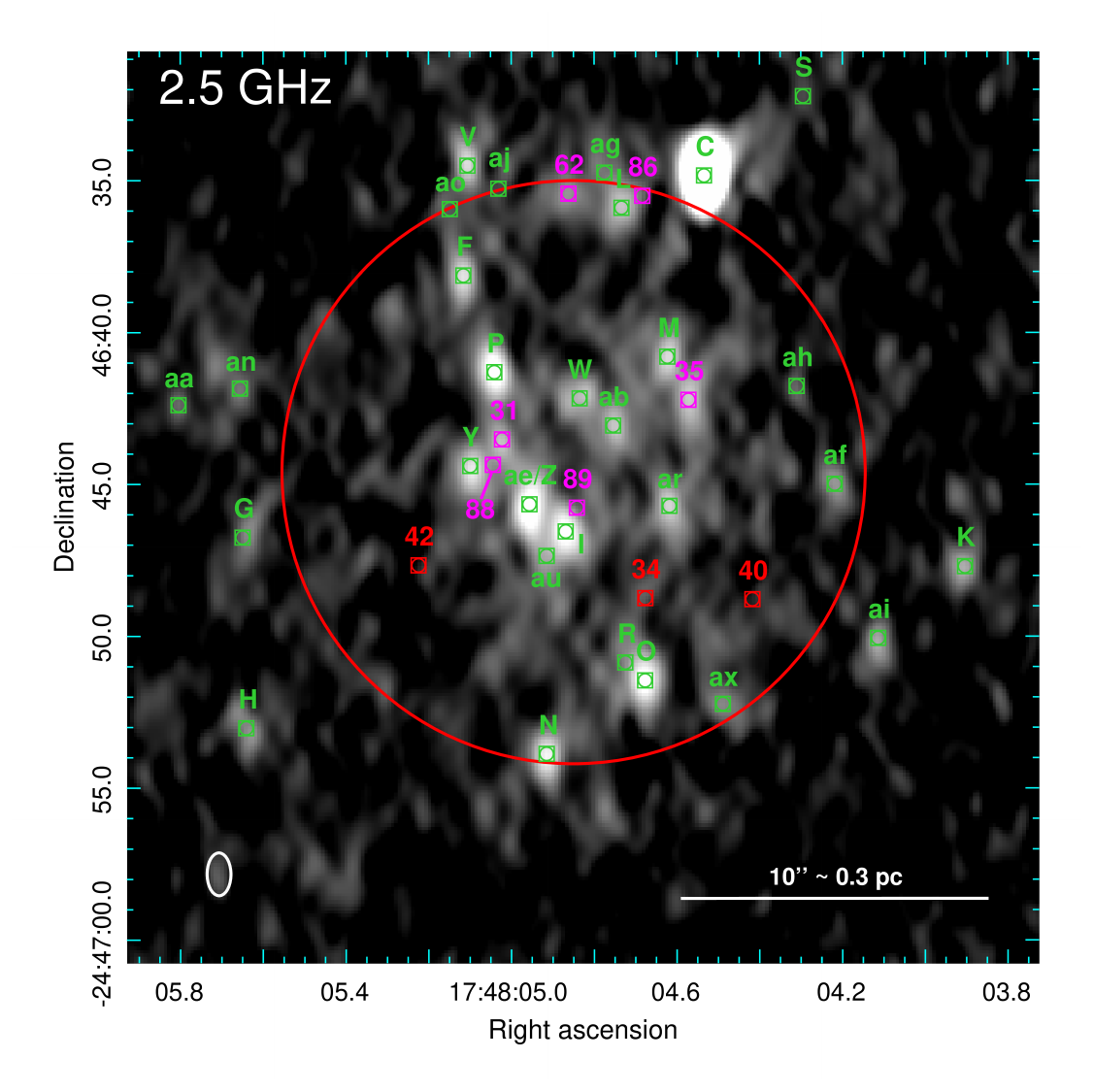}\\
    \includegraphics[width=0.495\textwidth]{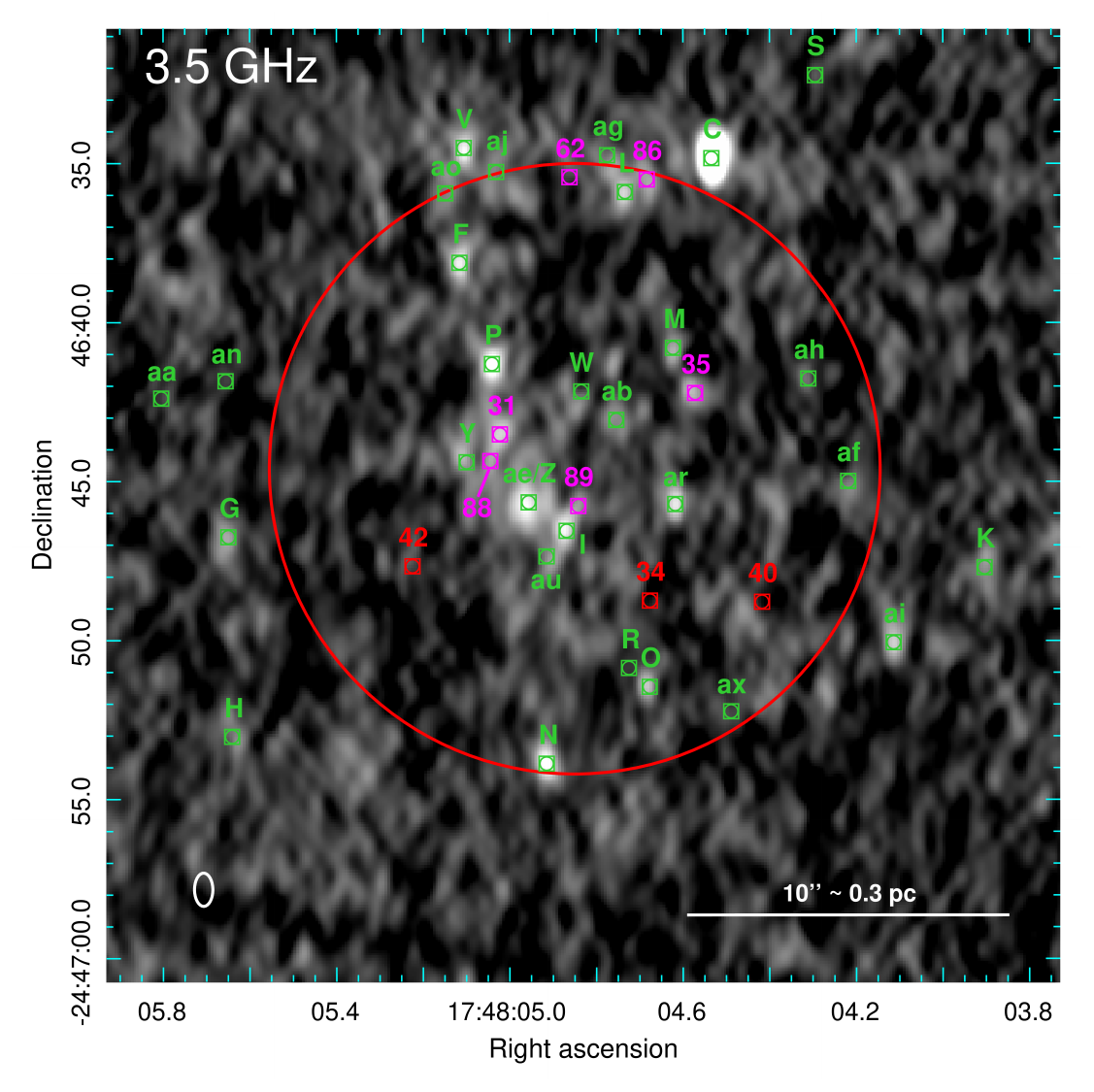}
    \caption{Top: 2.5\,GHz VLA image encompassing the core of Terzan\,5. The large red circle denotes the core radius of Terzan\,5. Green sources represent the continuum counterparts to timed millisecond pulsars and are labelled with their timing identification. Magenta sources are not associated with known timed pulsars. Red sources are from U20 but not detected in our current data. The FWHM synthesised beam size is indicated by the white ellipse in the bottom left corner. Bottom: same as in top panel, but for 3.5\,GHz.}
    \label{fig:image}
\end{figure}

\begin{figure}
    \centering
    \includegraphics[width=0.495\textwidth]{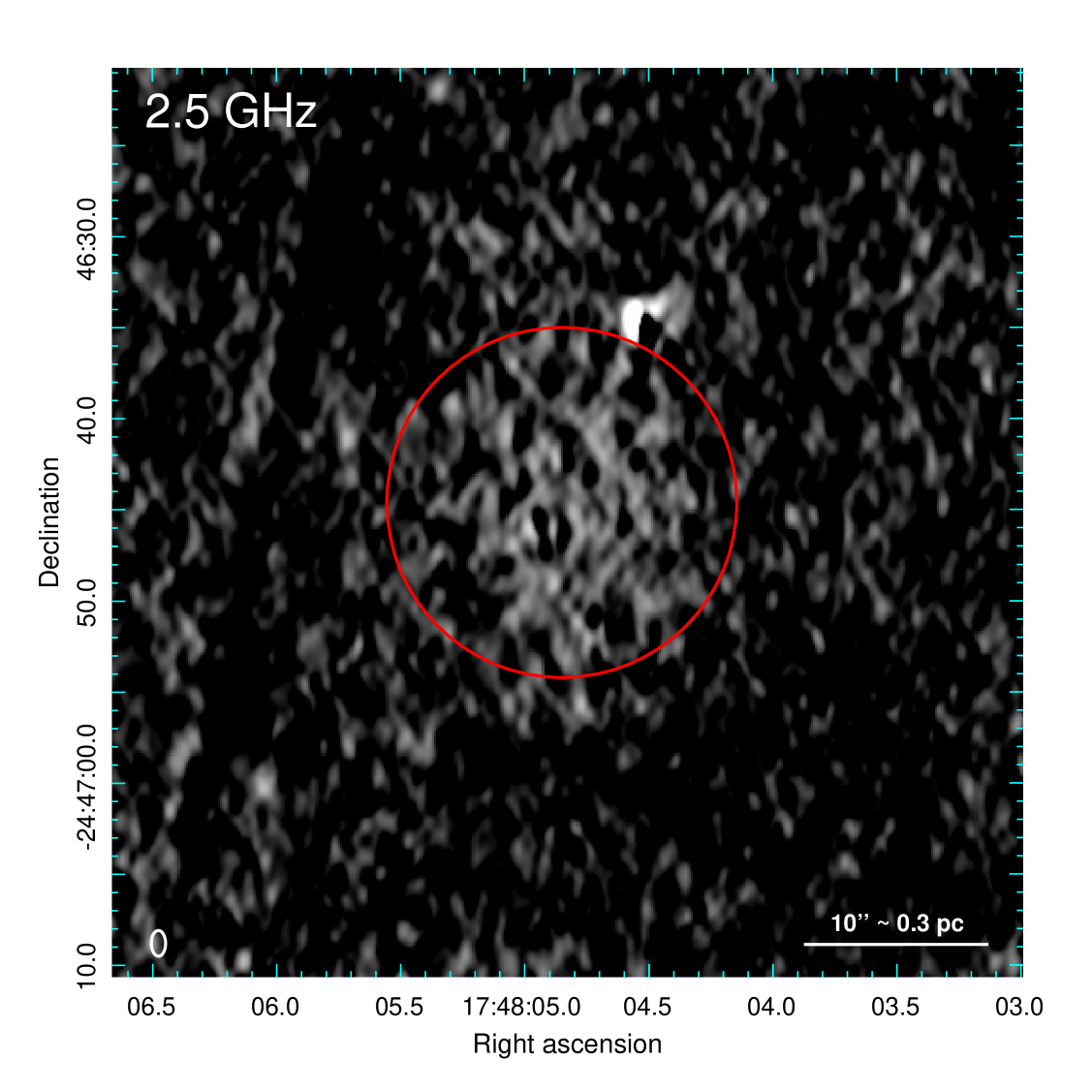}
    \caption{$50^{\prime\prime}\times50^{\prime\prime}$ residual 2.5 GHz image, with all significant point sources subtracted, of the central region of Terzan\,5; note the field of view is larger than in Figure \ref{fig:image}. The red circle shows the core radius. The white ellipse in the lower left corner indicates the beam size. Substantial unresolved emission remains within the core even after the subtraction of the significant point sources. The bright source to the north just outside the core is Ter5C, which has not subtracted fully cleanly due to its brightness and scintillation-induced variability.}
    \label{fig:residual}
\end{figure}

\section{Data Analysis} \label{sec:analysis}

Terzan\,5 was observed with the VLA across two 5.5\,hr blocks, one on 2022 March 14 (hereafter ``March'') and the other on 2022 April 16 (hereafter ``April''), for a total observing time of 11\,hr (Project ID: 22A-396). The data were taken at S band (2--4\,GHz) as the best compromise between sensitivity and spatial resolution, split into two 1.024\,GHz basebands centred at 2.5 and 3.5\,GHz. The VLA was in its most extended A configuration. 3C286 was used as the bandpass and flux calibrator. J1751--2524 was used as the phase calibrator.

We used the Common Astronomy Software Application (CASA; \citealt{2007ASPC..376..127M, 2022PASP..134k4501C}) to perform flagging, calibration and imaging. Data were processed using the VLA Calibration Pipeline, with additional manual flagging where needed. The two basebands were imaged separately, however, both 5.5\,hr observing blocks were concatenated together using CASA's \verb|tclean| task. We performed multi-frequency synthesis imaging; nterms=2 and a Briggs weighting with robust=1 were selected. 

Upon inspection of the image, we found significant artefacts as a result of the strong flux density variations of the bright eclipsing redback source Ter5A (see Section \ref{sec:ter5a} for the full details). We made time-resolved light curves, which were used to excise this source from both baseband images using the \verb|uvsub| command within CASA. Finally, each image was primary-beam corrected. We note that Terzan\,5 is not at the exact centre of the image: the phase centre is located $\sim1^{\prime}$ west of the core ($\approx90\%$ of the primary beam response at 2.5\,GHz and $\approx80\%$ at 3.5\,GHz).

The final images have a restoring beam of $1.41^{\prime\prime}\times0.79^{\prime\prime}$ (PA=0.3$^{\circ}$) and local rms of $\sim3\,$\fluxden for a central frequency of 2.5\,GHz and restoring beam of $1.05^{\prime\prime}\times0.58^{\prime\prime}$ (PA=0.2$^{\circ}$) and local rms $\sim2\,$\fluxden for 3.5\,GHz. They are shown in Figure \ref{fig:image}.

For consistency with U20, we assume the same values for the Terzan 5 core radius ($9.6^{\prime\prime}$; \citealt{Harris1996}) and distance (5.9 kpc; \citealt{Valenti2007}). Our analysis of the radio data is conducted using observed flux densities, so our inferences about the radio pulsar population do not depend on the specific distance assumed.

\subsection{Archival Radio Data Analysis}
\label{sec:arch}

To clarify the nature of the published diffuse flux measurement of the core of Terzan 5 from \citet{1990ApJ...365L..63F}, we reimaged their archival C-configuration radio data from 1989 Jun 12, adding in additional data taken with the same setup and configuration obtained on 1989 Sep 10 (project code TEST), BnC-configuration data from 1990 Oct 5 (project code AF206), D-configuration data from 1991 Apr 29 (project code TEST), and B-configuration data from 1990 Sep 7 (project code AF206). These data were all 
obtained in the continuum mode of the historic VLA, with two spectral windows each of 50 MHz bandwidth, centred at 1.465 and 1.515 GHz, for a mean frequency of 1.490 GHz. We reduced these data using standard routines in AIPS \citep{Greisen2003}, and concatenated all data together using {\tt DBCON}. Imaging with a Briggs robust value = 0 yields an image with $10.2^{\prime\prime} \times 7.3^{\prime\prime}$ (PA = $-1.7^{\circ}$) synthesised beam and an rms noise of 34.9 $\mu$Jy beam$^{-1}$. The resulting multi-configuration image has enhanced sensitivity to lower surface brightness emission and allows the bright pulsars Ter5A and Ter5C to be distinguished from the diffuse radio emission in the cluster core. We discuss the interpretation of this image in Section \ref{sec:diff}.

\subsection{Source finding and cross-matching catalogues}

For the new VLA A configuration images, we used the Python Blob Detection and Source Finder (PyBDSF; \citealt{2015ascl.soft02007M}) software package, version 1.9.0rc, to perform source finding. For consistency, we followed the same basic procedure described in U20. The primary-beam corrected 2.5\,GHz and 3.5\,GHz images were searched independently.
Within the half-light radius of Terzan\,5 ($r<0\bmins73$), we searched for, and selected, sources with a signal-to-noise (S/N) ratio of $>3$. Beyond the half-light radius we restrict our selection to more significant sources with S/N$>5$. As for previously published MAVERIC observations, we do not analyse sources beyond $3\bmins7$ from the cluster centre, equivalent to $\sim4$ half-light radii; objects beyond this radius are overwhelmingly background sources. All sources are listed in Table \ref{tab:src_list}; source ID numbers 1--43 match those of U20, with sources 44--89 being new sources not present in U20.

After this independent source finding, we force-fit for the fluxes at the timing positions of pulsars that were not yet detected, and retained the fluxes in our catalog only if the resulting detection was at least $2\sigma$. This led to flux density measurements for five additional known pulsars (Ter5S, Ter5T, Ter5aa, Ter5ac, Ter5ah), which are listed in Table \ref{tab:src_list}.

All flux densities are determined assuming objects are point sources: detected sources are fit with Gaussians fixed to the dimensions of the synthesised beam. This is appropriate for true Terzan 5 sources, all of which are expected to be point sources in these data, but will underestimate the flux of extended background sources, which are not our focus.

Owing to the background flux of unresolved pulsars in the core (Figure \ref{fig:residual}), the rms noise in this region is slightly higher than outside the core, leading to a brighter detection limit. We find $3\sigma$ detection limits of $>11.2 \mu$Jy at 2.5 GHz and $>7.8 \mu$Jy at 3.5 GHz in the core, which we take as our pulsar completeness limits.

\subsection{Astrometry and Cross-Frequency Matching}

To test the astrometry, we compare our new VLA positions of 18 bright, isolated pulsars to their precisely known positions determined by previous pulsar timing surveys. We identify a small but significant shift in Declination (Dec = -0\farcs14) and a negligible shift in Right Ascension (R.A. = 0\farcs008), both of which are applied to the VLA coordinates. The rms residuals between the two sets of positions are R.A.=0\farcs10 and Dec=0\farcs06, reassuringly consistent with the $1\sigma$ positional uncertainties estimated from the synthesised beam.

The initial positional uncertainties are determined by CASA {\tt imfit}. However, as suggested by the VLA Observational Status Summary\footnote{https://science.nrao.edu/facilities/vla/docs/manuals/oss/per formance/positional-accuracy}, we use 10\% of the full-width at half-maximum (FWHM) of the synthesised beam as a minimum value for the positional uncertainties. 

We cross-match the complete list of sources found in the two sub-band images. Sources are considered to match if their $3\sigma$ error ellipses overlap. The final catalogue includes 84 sources, with 37 associated with 38 known pulsars (Table \ref{tab:src_list}). We also recover the transitional millisecond pulsar candidate CX1 \citep{2018ApJ...864...28B}. The final International Celestial Reference System (ICRS) R.A. and Dec. of each source are calculated using the variance-weighted mean position. If a source is only detected in one of the subband images, we report the $3\sigma$ upper limit, calculated using the local noise measured from the PyBDSF rms images.

\subsection{Radio spectral analysis}

We calculate the spectral index $\alpha$ (defined as $S_{\nu} \propto \nu^{\alpha}$, where $\nu$ is the frequency and $S_{\nu}$ is the flux density at a given frequency) of each radio source. For non-detections, 3$\sigma$ upper limits are used. The spectral indices are modelled using the Bayesian Markov Chain Monte Carlo software \verb|JAGS| \citep{2012ascl.soft09002P} assuming a power law model and a uniform prior on $\alpha$ between $-$3.5 and 3.5. The modelling self-consistently takes into account both measurements (including uncertainties) and $3\sigma$ upper limits. The median of the posterior distribution of the spectral indices and 1$\sigma$ uncertainties are reported in Table \ref{tab:src_list}. For the sources only detected in one subband, a $3\sigma$ upper or lower limit on $\alpha$ is reported instead; these are prior-dependent.

\subsection{X-ray matching}

We follow the same procedure for finding X-ray counterparts to our radio sources that is outlined in U20. Where available, we use the X-ray properties from U20, otherwise we use the catalogue of \citet{goose}, which uses $\sim750$\,ks of archival \textit{Chandra}/ACIS observations. All X-ray matches were fit with an absorbed power-law model. Two absorption components were used; one fixed to the line-of-sight cluster absorbing column ($n_H=2.07\times10^{22}$\,cm$^2$), while the second is left free to account for intrinsic absorption. Sources with X-ray matches have their 0.5--10 keV X-ray luminosities listed in Table \ref{tab:src_list}.

\section{Results: Properties of Individual Sources}

\subsection{Comparison with U20}

\subsubsection{Flux Variations and Scintillation}
\label{sec:var}

\begin{figure}
    \centering
    \includegraphics[width=0.49\textwidth]{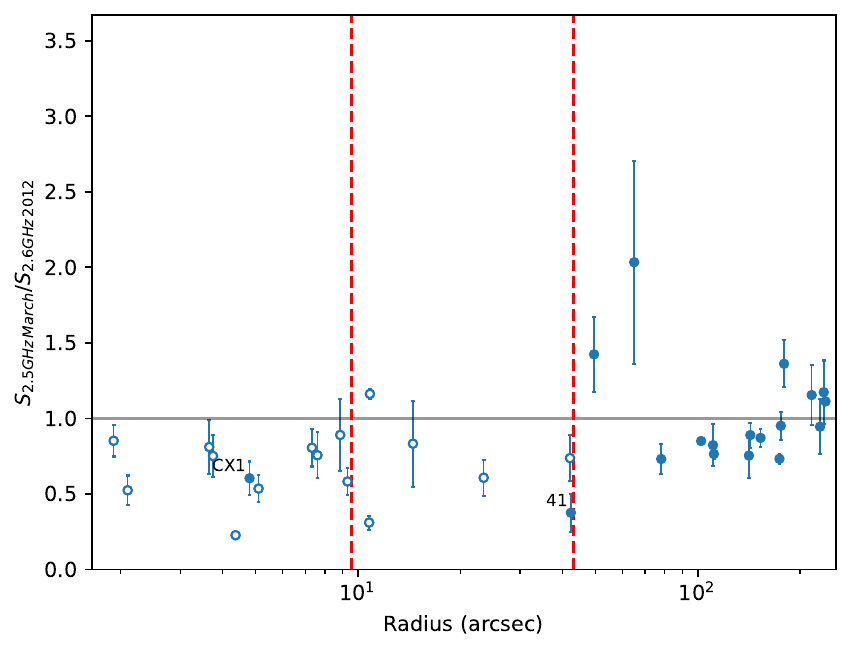}\\
    \includegraphics[width=0.49\textwidth]{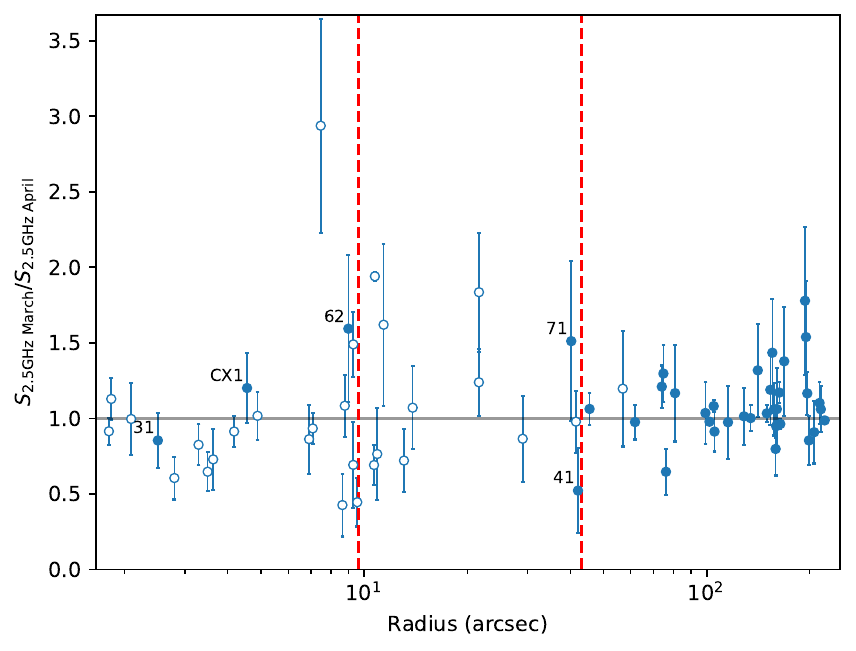}
    \caption{Top panel: Flux ratio between the 2.5\,GHz 2022 March 14 and the 2.6\,GHz 2012 observations for sources detected at both epochs. Unfilled circles are confirmed pulsars. There is some evidence that the 2022 flux densities for central sources are lower, likely due to scintillation. Bottom panel: Ratio of fluxes between March and April epochs. In both panels, red dotted lines represent the core and half-light radii. Sources within the half-light radius that are not confirmed pulsars are identified with their IDs.}
        \label{fig:var}
\end{figure}

The shallower catalogue of U20 contained 43 sources, close to half of the current catalogue. 38 of these are found in our new catalogue. In Figure \ref{fig:var}, we compare the 2.5 GHz flux densities of the sources detected at both epochs at this frequency as a function of radius. An immediate impression is that many of the sources within the half-light radius of the cluster are fainter in the new data compared to in U20, though this is only of modest statistical significance, with a median flux ratio of $0.74\pm0.23$. 

To the extent this is real, a possible explanation is refractive scintillation, expected to have an rms value of 33\% at 2.5 GHz (using the methodology of \citealt{Hancock2019}). Hence the amplitude of the observed flux variations is consistent with those expected from scintillation. There is existing evidence that the \citet{Hancock2019} model does a good job of reproducing the Ter 5 scintillation properties: it predicts an rms variation of 26\% at 1.75 GHz, compared to an observed value of $25\pm4$\% \citep{Martsen2022}. The sources at large projected radii do not share the apparent behaviour of the central sources, but this is not straightforward to interpret, as these are likely background sources that may have larger angular sizes that would not be expected to show scintillation.

A caveat to the interpretation of refractive scintillation for the central flux variations is that \citet{Martsen2022} did not find clear spatial correlations among the pulsar flux variations, the opposite of the apparently correlated variations we observe. Unfortunately, there is no overlap in time between their published observations and our new continuum data that would allow a direct comparison of the flux densities at a similar time.

The timescale of refractive scintillation is harder to predict as it depends on the relative screen velocity, which is a priori unknown. Assuming a value of 100 km s$^{-1}$, it is 4.6 d at 2.5 GHz. Hence, in principle, flux variations could also be detected between the March and April epochs. We compare these in the bottom panel of Figure \ref{fig:var}. There is no clear evidence for variation on this $\sim$ month-long timescale, perhaps broadly consistent with the idea that the time difference corresponds to several refractive timescales, but these observations are not very constraining. 

The consistency of the March and April flux densities at all radii, as well as the large-radius consistency between 2012 and 2022, together suggest that there is no issue with the flux density scale of the 2022 March observations. Future time series observations of Terzan 5 would be valuable to compare to predictions of scintillation models and see whether there is additional evidence of spatially correlated flux variations.

\subsubsection{Missing Sources}

Five of the sources from U20 are not present in the new catalogue. The non-detection of Ter5-VLA42 is expected---it is the transient low-mass X-ray binary EXO 1745--248, which was only detected at the higher frequencies in the previous data. Ter5-VLA21 and Ter5-VLA25 are located far outside the half-light radius and had radio spectral indices $\alpha = -0.7$ to --0.9: they are almost certainly background sources unassociated with the cluster. 

The remaining two sources, Ter5-VLA34 and Ter5-VLA40, were both located inside the core radius and hence are very likely to be associated with the cluster. On the basis of its X-ray luminosity ($L_X \sim 8 \times 10^{31}$ erg s$^{-1}$; 0.5--10 keV), hard X-ray photon index, and poorly characterised radio spectral slope, U20 classified Ter5-VLA34 as a likely redback millisecond pulsar. While it is formally not detected in our new data, there is some flux present at this position in the 2.5 GHz image, with an approximate value of $\sim 8\mu$Jy; there is no flux at 3.5 GHz. In the newer X-ray analysis of \citet{Kumawat2025} the source has a similar X-ray luminosity and is still found to have a hard power law, with $\Gamma = 1.1\pm0.5$. It seems plausible that the identification of Ter5-VLA34 as a spider is accurate but that it is potentially eclipsed in the new data, leading to it being fainter in the radio. In this case it may be discovered by future pulsar search observations. Alternatively, the X-ray source could be an accreting white dwarf, and the radio source an unrelated faint pulsar.

The case for Ter5-VLA40 is somewhat similar to Ter5-VLA34, in that U20 identified it as a likely redback pulsar and it is not present in the new catalogue. But there are some differences: its radio spectral index was well-measured in U20 and consistent with a pulsar ($\alpha = -2.7^{+0.8}_{-0.5}$), while its X-ray luminosity is lower ($L_X \sim [1$--$2] \times 10^{31}$ erg s$^{-1}$). Hence there is undoubtedly a yet-untimed pulsar at this location. If it is a spider, its non-detection in the new data could be explained as an eclipse, though as for Ter5-VLA34, it could also be a chance alignment between a non-spider pulsar and an unrelated X-ray source.

The U20 locations of Ter5-VLA34, Ter5-VLA40, and Ter5-VLA42 are plotted in Figure \ref{fig:image}.

\subsubsection{Ter5-VLA31: Candidate accreting black hole} \label{sec:BH_can}

Ter5-VLA31 was previously identified in U20 as an accreting stellar-mass black hole candidate due to its flat radio spectrum ($\alpha = +0.2\pm0.2$), radio/X-ray flux ratio, with an X-ray counterpart of $L_{x} =(2.1\pm0.5) \times 10^{31}$ \ergs, and its location within the core. 

In our new radio data, we find some evidence for a steeper radio spectral index 
($\alpha=-1.4^{+0.6}_{-0.7}$), though with much larger uncertainties than before, owing to the lack of higher frequency data. The actual flux density estimates are not formally inconsistent with the 2012 measurements: at 2.5 GHz we now find $27\pm4\mu$Jy versus $<20 \mu$Jy before, and at 3.5 GHz the measurements agree: $17\pm3\mu$Jy compared to $22\pm4\mu$Jy before. Hence the evidence that there is a change at all is not strong but only suggestive.

To the extent that the measured properties of the source have changed, one possibility is that a subset of the measurements are affected by source confusion in the core, which could vary between the epochs due to scintillation. We note that there is a new timed pulsar (Ter5av; \citealt{Padmanabh2024}) located near the position of Ter5-VLA31. A priori we do not expect to detect Ter5av in our data: its expected flux density is $\sim 5\mu$Jy at 2.5 GHz.

Examining the 2012 images at all frequencies from U20, VLA31 does not necessarily have a consistent position among them. At 2.6 GHz it cannot be confidently separated from the much brighter Ter5P. At 3.4 GHz there is a high-confidence source potentially consistent with Ter5av (separated by $\sim 0.25-0.3\arcsec$), while the 5.0 GHz image shows a well-detected ($32\pm3\mu$Jy) source that definitely does not match Ter5av (separated by $0.96\arcsec$).

\citet{Padmanabh2024} also suggest the possible association of the X-ray source that we associate with Ter5-VLA31 with Ter5av, but this pulsar is not in a short period orbit and shows no eclipses, so does not appear to be a spider pulsar that would be expected to have X-ray emission at the level observed.

One plausible explanation of these data is that there are two nearby sources, Ter5av and a flatter-spectrum source, and our measured flux values for Ter5-VLA31 are a mix of these that vary with frequency, resolution, and time. It could instead be true that Ter5av is too faint to contribute to most of our data. In either case, the identification of the flatter-spectrum source as a candidate black hole is still plausible.

We can rule out an alternative explanation of Ter5-VLA31 as a transitional millisecond pulsar as its X-ray luminosity is much lower than the expected accretion state value of $L_X \sim 10^{33}$--$10^{34}$ \ergs (e.g., \citealt{Linares2014}), and  there is no evidence for X-ray variability over a large number of X-ray observations spanning 13 years \citep{Kumawat2025}, including annual observations from 2011 to 2014. 

Future deep C or X band data would be ideal to reduce source confusion around this source by minimising pulsar emission, revealing the flat-spectrum emission if present.

\subsection{Ter5A: The luminous eclipsing pulsar}
\label{sec:ter5a}

Ter5A is the brightest radio continuum source in the cluster (it is at a projected radius of 0.6\arcmin\ from the centre and hence outside the field of view of Figure \ref{fig:image}). First discovered by \citet{Lyne1990}, it is an eclipsing redback millisecond pulsar with orbital period of 1.8 hr. Previous studies of the eclipses have led to the conclusion that the predominant eclipse mechanism is free--free absorption from the ionised wind from the low-mass companion star \citep{Rasio1991,Thorsett1991}.

\begin{figure}
    \centering
    \includegraphics[width=0.5\textwidth]{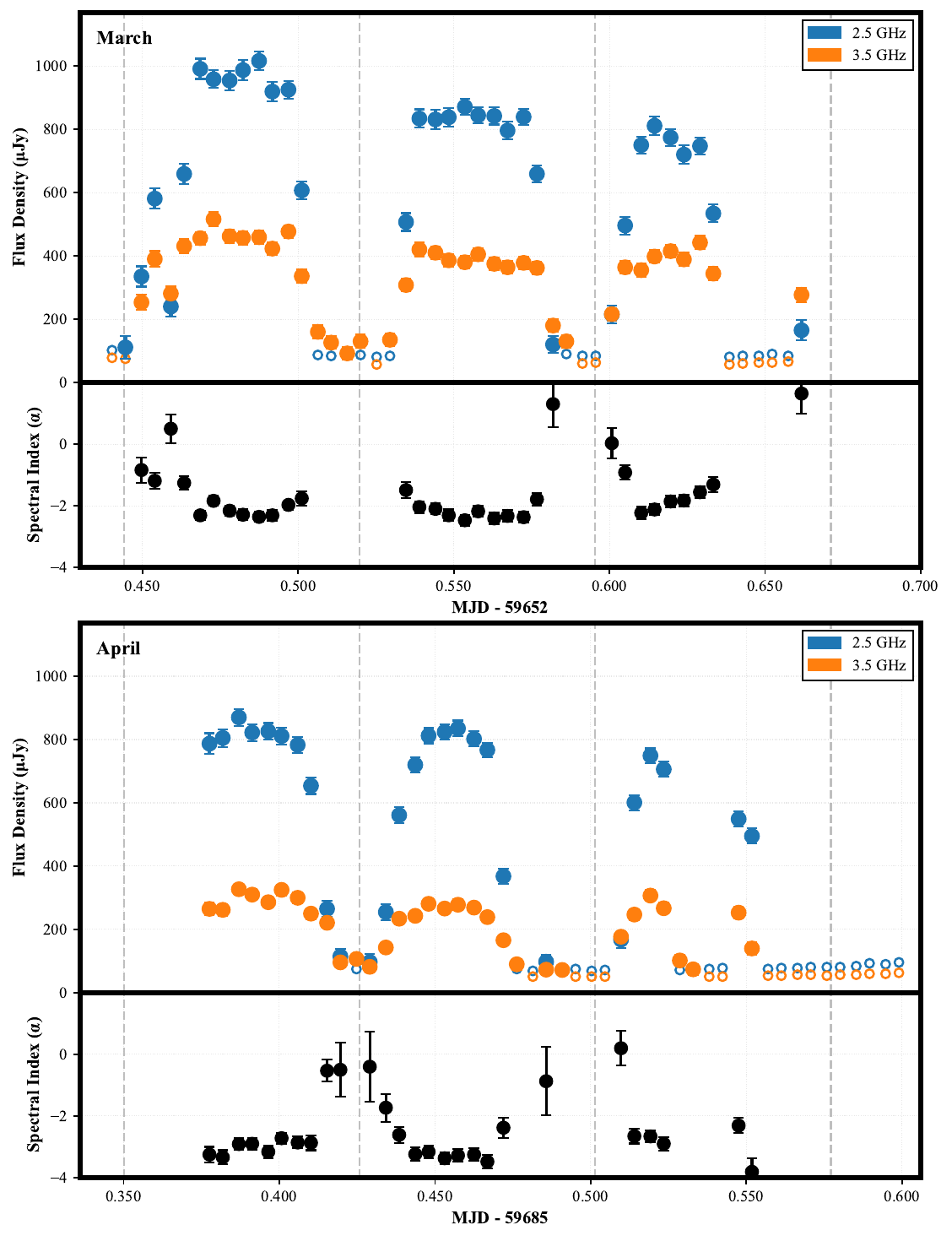}
    \caption{March (top) and April (bottom) flux density and spectral index measurements for the eclipsing redback Ter5A, with filled circles detections and unfilled circles $3\sigma$ upper limits. The dashed lines represent the expected conjunction times for the companion to be in front of the pulsar. The length, depth, and timing of eclipses varies substantially in these data. Data used to created each light curve are reported in Table \ref{tab:ter5a_flx}.}
    \label{fig:lc_ter5A}
\end{figure}

Each of the two 5.5-hr execution blocks of our VLA observations covers close to 3 orbital periods of Ter5A. We imaged Ter5A in each individual scan in both blocks, leading to a light curve with typical sampling of $\sim 6$--7 min for each block in both 2.5 and 3.5 GHz. These light curves are listed in Table \ref{tab:ter5a_flx} and shown in Figure \ref{fig:lc_ter5A}. In each panel the inferior conjunctions of the companion are also marked, using the Ter5A ephemeris from \citet{Rosenthal2025}. These are the epochs where the absorbing material might naively be expected to be thickest.

On 2022 March 14, the light curve captures the egress of one eclipse, two full eclipses, and most of a fourth. In each case the pulsar is not detected at 2.5 GHz for a varying length of time, ranging from $40$ min for the first full eclipse to only 20 min for the second full eclipse. We have taken these eclipse lengths as the time from the start of the first fully eclipsed scan to the end of the last fully eclipsed scan; they are necessarily quantised by the scan lengths. The timing of the eclipses is also not stable: the first full eclipse is centred at conjunction, while the second and especially the last eclipses start ``early'', suggesting the absorbing material is concentrated ahead of the companion in its orbit, rather than trailing it.

In addition to this variation in eclipse lengths, the behaviour at 3.5 GHz also varies among the eclipses. In the first full eclipse, it is detected in all but one scan, at an average flux value $\sim 35$\% of the preceding out-of-eclipse flux. But it is undetected for the last two scans of the shorter second full eclipse, and only detected in a single scan in the last eclipse.

Variations in the eclipse properties are also seen in the 2022 April 16 light curve (Figure \ref{fig:lc_ter5A}, bottom). This second light curve also shows a ``bonus'' eclipse occurring around superior conjunction of the companion, only 30 min after the previous eclipse ends, as well as an unusually long complete eclipse at both frequencies, lasting at least 1.1 hr, at the end of the data set.

In March the spectral indices show that absorption is present at the majority of epochs even outside of the obvious eclipses: as the 2.5 GHz flux is more absorbed than the 3.5 GHz flux, the spectral index becomes less steep or in a few cases is even inverted such that the pulsar is brighter at higher frequency. In April there is less evidence for absorption outside of eclipses.

Eclipses of varying length and sometimes at unexpected phases have been previously observed for Ter5A in continuum at lower frequencies (e.g., \citealt{Thorsett1991,Smirnov2025}). These new data show multiple eclipses of Ter5A in continuum at the relatively high frequency of 3.5 GHz, though we note that another redback, PSR J1740--5340 in the globular cluster NGC 6397, has shown continuum eclipses at an even higher frequency of 5.5 GHz \citep{Zhao2020}. Our high signal-to-noise multi-band measurements of eclipses in Ter5A should prove useful in constraining the physical parameters of the eclipsing material in a future work.

\subsection{Properties of Known Pulsars}

There are currently 49 pulsars in Terzan 5 that have been confirmed via pulsar timing \citep{Padmanabh2024}, of which 48 (all but Ter5al) have timing positions. Of these 48 pulsars, 38 are present in our point source catalogue as 37 separate sources: the pulsars Ter5Z and Ter5ae are too close to be separated in our data, and are detected together as a single source. 

While few of the timed pulsars have been observed at 2.5 GHz, we can predict their 2.5 GHz fluxes from timing fluxes calculated from the radiometer equation. The most reliable of these are probably the long-term 2.0 GHz fluxes and spectral indices for 31 pulsars listed in \citet{Martsen2022}, but for the 17 pulsars not included in that paper, we take the 1.3, 1.4, or 2.0 GHz flux density from the relevant papers \citep{Ransom2005,Hessels2006,Andersen2018,Cadelano2018,Ridolfi2021,Padmanabh2024} and assume $\alpha=-1.8$ to calculate the expected 2.5 GHz flux. For the two pulsars not in \citet{Martsen2022} but with relatively well-measured spectral indices in our continuum data (Ter5P and Ter5ar), the values are consistent with 
$\alpha=-1.8$, suggesting this is a reasonable assumption.

\begin{figure}
    \centering
    \includegraphics[width=0.9\linewidth]{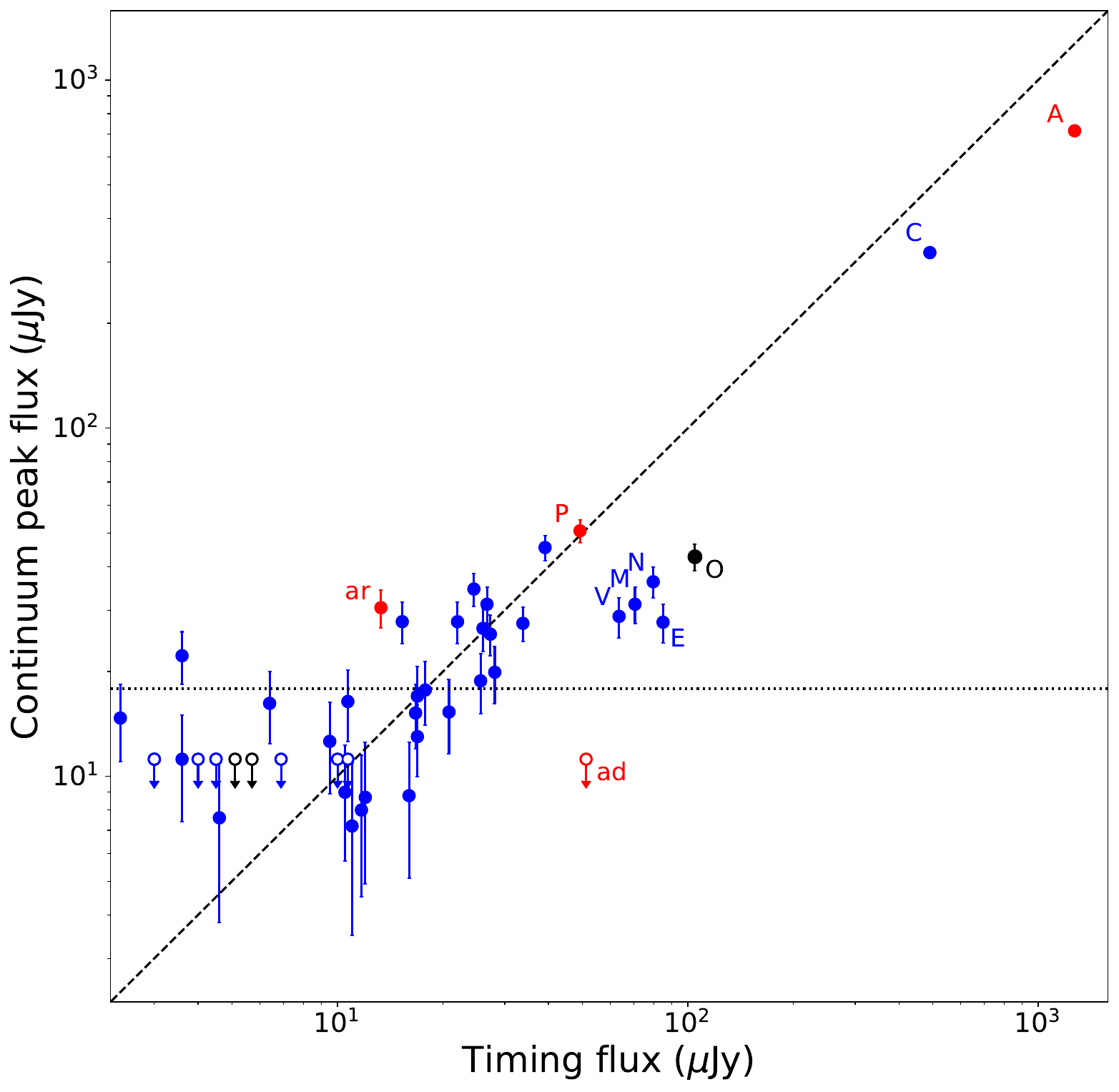}
    \caption{2.5 GHz VLA flux densities of known Terzan 5 pulsars plotted against their timing fluxes. Timed pulsars are converted to 2.5\,GHz using the spectral index from \citet{Martsen2022} for the 31 pulsars where this information is available. For the remaining pulsars that do not have timing spectral information, we use an assumed $\alpha=-1.8$. Upper limits are denoted  with unfilled circles. Red and black points indicate eclipsing redback (Ter5A, Ter5P, Ter5ad, Ter5ar) and black widow (Ter5O, Ter5aq, Ter5at) sources, respectively. The diagonal dashed line represents unity. The horizontal dotted line indicates our new candidate pulsar Ter5-VLA62. We label notable pulsars (e.g., redbacks, bright sources and outliers).}
    \label{fig:fluxes_timed_vs_continuum}
\end{figure}

This comparison is given in Figure \ref{fig:fluxes_timed_vs_continuum}. 
Similar to the comparison between 2012 and 2022 continuum fluxes in Figure \ref{fig:var}, we see that the 2022 continuum pulsars are mostly fainter than predicted by their timing flux. For Ter5A, Ter5O, and Ter5P this could be due to eclipses, but that is not the case for the the non-spider pulsars. Since there is not a similar offset between the U20 2012 continuum fluxes and the timing-predicted fluxes, scintillation in our data is likely the main culprit (Section \ref{sec:var}). However, it may not be the only mechanism at play. \citet{Martsen2022} find that many of these pulsars have flatter spectra than typical, with a mean spectral index of $\alpha = -1.35$. They attribute this to a selection effect from observations at 2.0 GHz rather than the lower frequencies typical of pulsar searches. There are at least two pulsars in our data (Ter5M and Ter5Y) that we find to have much steeper spectral indices than in \citet{Martsen2022} at high confidence. 
It will be worth seeing whether such differences persist in future datasets, or whether they are due to scintillation-induced variations in our measured spectral indices.

Of the ten undetected pulsars, nine (Ter5U, Ter5ak, Ter5am, Ter5ap, Ter5aq, Ter5as, Ter5at, Ter5av, Ter5aw) have predicted 2.5 GHz flux densities from $\sim 3$--$10 \mu$Jy, so their non-detections are expected.

The last undetected pulsar is the redback Ter5ad, which was also not detected in U20. Given its reported 1.95 GHz flux density of $80\mu$Jy \citep{Hessels2006}, for a typical spectral index this would imply a flux density of $\sim 50\mu$Jy at 2.5 GHz and $\sim30\mu$Jy at 3.5 GHz, where it should be easily detected in our new data. While we cannot rule out the possibility that it is eclipsed in all our datasets (both old and new),
it is plausible that its intrinsic flux density is fainter than the value reported in \citet{Hessels2006}, and a combination of intrinsic faintness and eclipses are preventing its detection.

All the detected pulsars in the centre of the cluster are labelled in Figure \ref{fig:image}.

\subsection{Unassociated core sources}
\label{sec:unass}

In Table \ref{tab:src_list}, there are seven continuum sources in the core without a clear association. Of these seven, the candidate spider pulsars Ter5-VLA34 and Ter5-VLA40 from U20 were discussed above, as was the candidate black hole Ter5-VLA31.

Of the remaining four sources, Ter5-VLA62 is the only unassociated steep-spectrum source within the core and is a new candidate millisecond pulsar. Ter5-VLA62 is significantly detected in our lower subband (2.5\,GHz; $18\pm4\mu$Jy) and is not coincident with a known pulsar. This source was not detected in U20, and does not have an X-ray counterpart.

Ter5-VLA86, Ter5-VLA88, and Ter5-VLA89 are all detected only at 3.5 GHz, and were not detected by U20. Each source lies in a crowded region of the core and at 2.5 GHz appears to be contaminated by emission from neighbouring bright pulsars, which could hide real 2.5 GHz flux. Hence we cannot rule out the possibility that these sources are indeed pulsars. Of these, the only one with a potential X-ray match is Ter5-VLA89, located $0.7\arcsec$ from the X-ray source CX9. This X-ray source  is classified as a quiescent neutron star low-mass X-ray binary \citep{Kumawat2025}. Such binaries are not expected to show radio emission and the offset is large enough that these sources are probably unrelated.

\section{Results: Pulsar Population of Terzan\,5} \label{sec:results}

Given the large pulsar population of Terzan 5 and the detection of radio continuum sources at our detection limit in the 2012 data, the central scientific goal of the new observations was to better characterise the population of pulsars by increasing both the depth and spatial resolution of the VLA imaging.

\subsection{Pulsar Luminosity Function}

We construct an empirical pulsar luminosity function from our continuum data, starting from the 37 continuum sources present in our point source catalog. We split the combined flux density of the overlapping pulsars Ter5ae and Ter5Z based on their relative 1.4 GHz flux densities \citep{Martsen2022}, assigning a flux density of $36.6\pm3.8 \mu$Jy to Ter5ae and $19.6\pm3.8 \mu$Jy to Ter5Z. This gives 38 confirmed pulsars with 2.5 GHz continuum flux estimates.
 
 As highlighted above in Section \ref{sec:unass}, Ter5-VLA62 has not yet been detected as a timed pulsar, but has a steep radio spectrum and is located in the cluster core. We consider Ter5-VLA62 to be high-probability pulsar and include it in our luminosity function, giving 39 pulsars. Two additional pulsar candidates identified in U20 (Ter5-VLA34 and VLA40) are not detected in the new imaging; these are plausibly real pulsars that are fainter in these data due to scintillation, but we do not include them here.

\begin{figure}
    \centering
    \includegraphics[width=0.49\textwidth]{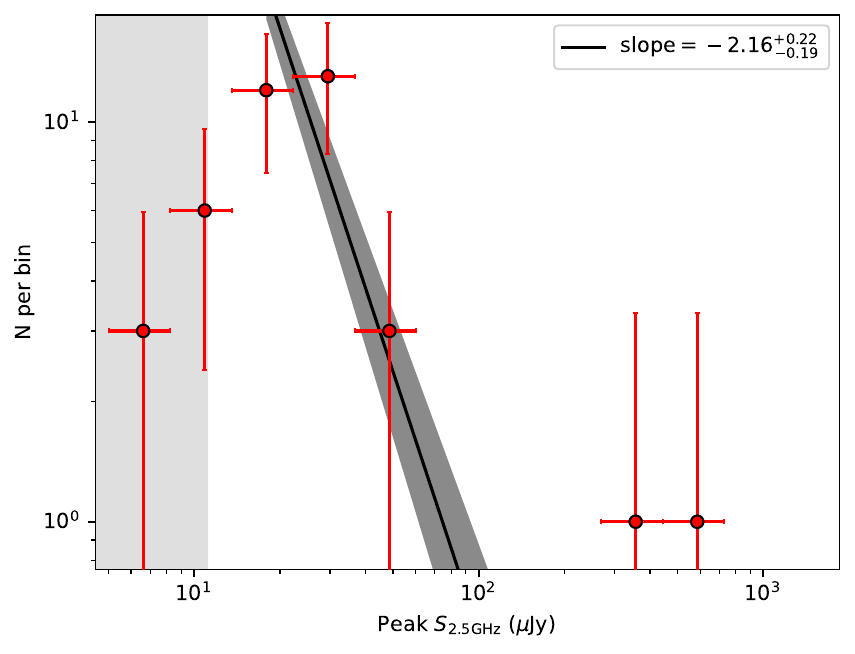}
    \caption{2.5 GHz luminosity function for the pulsars in Terzan\,5. The black line shows the best fit spectral index with grey $1\sigma$ uncertainty ($\alpha_{p}=2.16^{+0.22}_{-0.19}$). The vertical grey shaded region indicates where we are incomplete ($<11.2\,\mu$Jy). The obvious turnover in the luminosity function is due to this incompleteness, though there is a hint of a change in slope in the more complete region.}
    \label{fig:lum_fun}
\end{figure}

Figure \ref{fig:lum_fun} shows the 2.5 GHz continuum luminosity function of the 39 high-probability pulsars, with a power-law fit to the complete ($>11.2\mu$Jy) sample overplotted. The superficial turnover in the data at fainter fluxes is due to incompleteness, though there is a hint of a change in slope around $\sim 20$--$30 \mu$Jy, above the completeness limit of the sample. 

To further investigate the pulsar population, we next calculate the total flux of pulsars in the cluster, including unresolved flux.

\subsection{Unresolved Flux Measurement}
\label{sec:diff}

\subsubsection{New Data}

Figure \ref{fig:residual} shows that there is well-detected ``residual" flux from the pulsars that are fainter that the detection limit of our new A configuration observations. This flux is apparent because in many beams there are a sufficient number of sub-threshold pulsars whose summed flux is detectable. We can constrain the total pulsar population in Terzan 5 by estimating the integrated amount of flux in these subthreshold pulsars.

Owing to the limited sensitivity of our A configuration observations to low surface brightness emission, we are not sensitive to the flux from subthreshold pulsars located at large distances from the centre. Therefore, to calculate the \emph{total} diffuse flux, we have to choose a radius in which we have a high signal-to-noise measurement of the diffuse flux and then extrapolate to get an integrated measurement over all radii.

To do this, we assume the spatial distribution of the faint pulsars producing the diffuse flux follows the brighter resolved pulsars. To be self-consistent, we only consider the radial distribution of the \emph{detected} pulsars in our data. Of the 39 pulsars, we find that about half (20/39; 51.2\%) are found within a radius of 9.6\arcsec\ (the core radius), and about three-quarters (29/39; 74.4\%) are found within a radius of 13.3\arcsec. We integrate the unresolved flux in our data out to these radii, and extrapolate based on the above fractions to estimate the total unresolved flux integrated over all radii. The uncertainties in these quantities include both thermal noise and the uncertainty in the extrapolation. Separately, we add up the resolved flux from pulsars out to these radii, but when calculating the total integrated resolved flux over all radii, we use the actual (not extrapolated) measurements, since pulsars above our detection limit should be detectable at all radii within our field of view. Table \ref{tab:residual_flux} shows the integrated resolved (from detected sources listed in our catalogue) and unresolved flux within these radii, as well as an extrapolated value integrated over all radii. The unresolved flux is bright: is is comparable to or larger than the summed flux of resolved pulsars (excluding Ter5A and Ter5C). We get consistent values from both radius choices, in the range 1.8--2.0 mJy, consistent to within the listed uncertainties in the quantities.

\begin{table}
    \centering
    \renewcommand{\arraystretch}{1.5}
    \caption{Integrated 2.5 GHz Fluxes from New VLA Data}    
    \resizebox{\linewidth}{!}{\begin{tabular}{lccccc}\hline\hline
Radius & $F_{src}^{\mathrm a}$ & $F_{resid}^{\mathrm b}$ & $F_{tot}^{\mathrm c}$ & $F_{tot,ext}^{\mathrm d}$ & $F_{tot,ext,A/C}^{\mathrm e}$\\
       & ($\mu$Jy)  & ($\mu$Jy)    & $(\mu$Jy)     & ($\mu$Jy) & ($\mu$Jy) \\ \hline
   $< 9.6\arcsec^{\mathrm f}$   & $548\pm16$ & $578\pm45$ & $1126\pm48$ & $1955\pm198$ & $2988\pm225$ \\
   $< 13.3\arcsec^{\mathrm g}$  & $677\pm20$ & $723\pm63$ & $1400\pm66$ & $1800\pm127$ & $2833\pm166$ \\\hline
    \end{tabular}}
\begin{flushleft}
$^{\mathrm a}$\hspace{0.1cm} Summed 2.5 GHz flux from the resolved pulsars in our catalogue within this radius, excluding Ter5C from $< 13.3\arcsec$. \\
$^{\mathrm b}$\hspace{0.1cm} Integrated unresolved 2.5 GHz flux within this radius.\\
$^{\mathrm c}$\hspace{0.1cm} Total (resolved+unresolved) 2.5 GHz flux flux within this radius, excluding Ter5C from $< 13.3\arcsec$. \\
$^{\mathrm d}$\hspace{0.1cm} Total (resolved+extrapolated unresolved)  2.5 GHz flux attributed to pulsars in Terzan 5, excluding Ter5A/Ter5C.\\
$^{\mathrm e}$\hspace{0.1cm} Total (resolved+extrapolated unresolved)  2.5 GHz flux attributed to pulsars in Terzan 5, including Ter5A/Ter5C. We have assumed a 15\% uncertainty in the Ter5A flux density, which roughly reflects out-of-eclipse flux variations but not the deepest eclipses. \\
$^{\mathrm f}$\hspace{0.1cm} This radius includes 20/39 (51.3\%) of the pulsars detected in the new VLA data.\\
$^{\mathrm g}$\hspace{0.1cm} This radius includes 29/39 (74.4\%) of the pulsars detected in the new VLA data.
\end{flushleft}
\label{tab:residual_flux}
\end{table}

Our new measurements are the first that allow simultaneous measurements of the resolved and diffuse flux at the same frequency and hence are well-suited for modelling the pulsar population.

We acknowledge several additional sources of uncertainty in our new measurements. First, our 2.5 GHz residual image (Figure \ref{fig:residual}) shows imperfect subtraction of the point sources (visible for Ter5C, right outside the core). To assess how this might affect our diffuse flux measurement, we compared the summed diffuse + resolved source flux measurement to the integrated flux out to 9.6\arcsec\ and 13.3\arcsec\ in the original unsubtracted 2.5 GHz image. We find agreement within 4\% for 9.6\arcsec\ and 1\% for 13.3\arcsec. The differences are likely a combination of the imperfect subtraction of point sources as well as edge effects: several sources cross these radial limits and are only partially counted in a direct pixel integration. These small differences are subdominant to the other uncertainties in our measurement so do not meaningfully affect our analysis.

A second source of uncertainty is that A-configuration images lack sensitivity to low surface brightness emission, so it is difficult to be sure that we are sensitive to ``all" the emission within our chosen radii. To address this potential concern, we next compare our results with those  from archival lower angular resolution data. To anticipate these results, we find no evidence that we are missing substantial diffuse flux, at least within the core.

\subsubsection{Comparison To Archival Data}

The most direct comparison is to the results of \citet{2000ApJ...536..865F}, who reported that their 1.49 GHz A-configuration images had $\sim 2$ mJy ``central" flux (excluding Ter5A and Ter5C) that was diffuse compared to their beam of $5.9\arcsec \times 4.1\arcsec$. It is difficult to precisely define the outer edge of the  \citet{2000ApJ...536..865F} diffuse flux measurement, but it does not appear to extend beyond the $10.7\arcsec$ projected radius of Ter5C. When combined with the spectral index they measured for the diffuse flux of $\alpha = -1.65\pm0.16$, this corresponds to a predicted ``central" flux of $\sim 850 \mu$Jy at the 2.5 GHz central frequency of our new observations. This value is about 25\% smaller than, but generally consistent with, the measurement of 1.1 mJy that we find within the cluster core, especially given that the outer boundary of their measurement is not well-defined.

\begin{figure}
    \centering
    \includegraphics[width=0.495\textwidth]{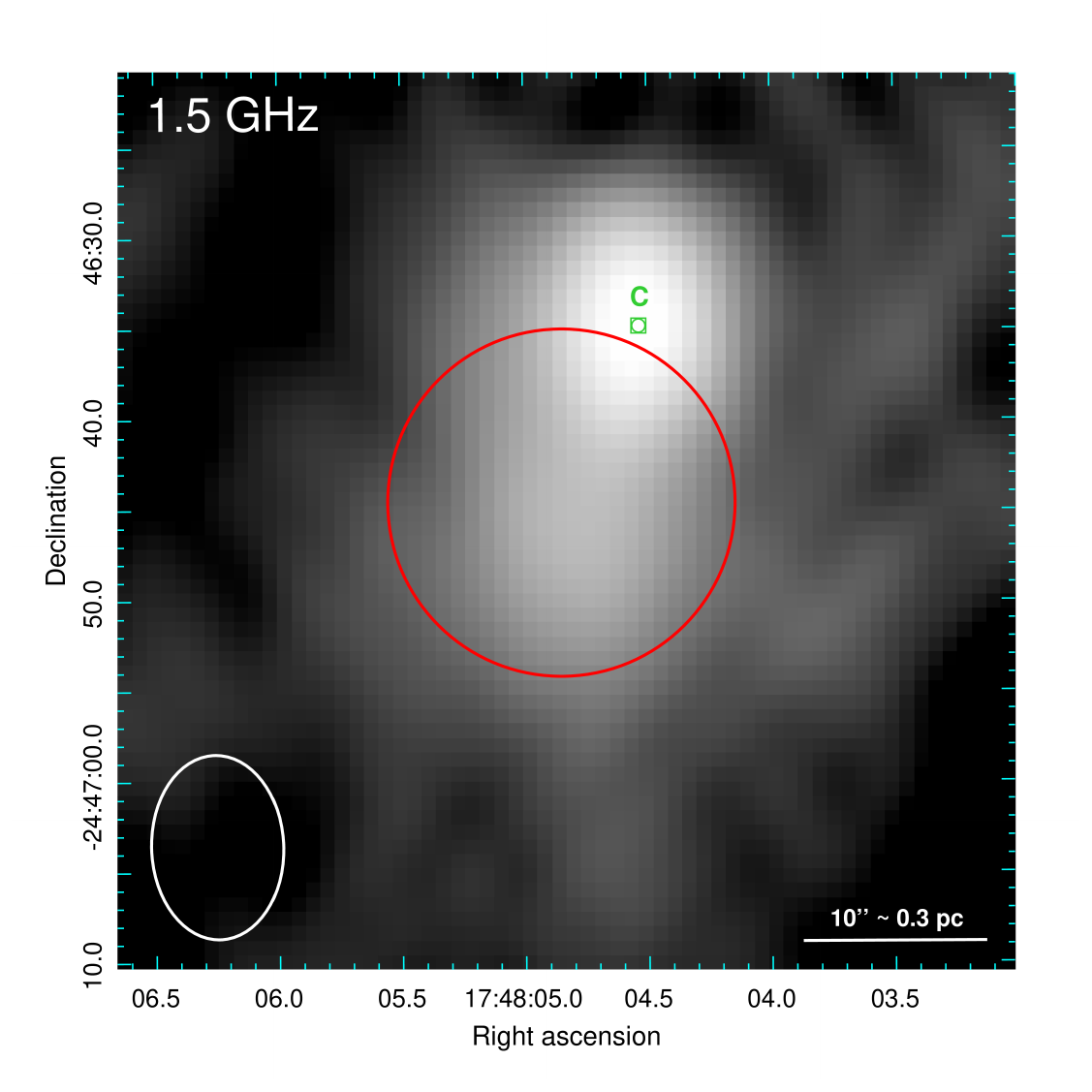}
    \caption{1.5\,GHz pre-upgrade VLA multi-configuration (B+C+D) image of Terzan\,5, covering the same area as the newer images in Figure \ref{fig:image}. The large red circle denotes the core of Terzan\,5. The FWHM synthesised beam size is indicated by the white ellipse in the bottom left corner. Due to the lower resolution compared to the newer A configuration data shown in Figure \ref{fig:image}, Ter5-C is the only resolved source within the image. Nonetheless, the diffuse core emission in this image appears consistent with that measured in our new high-resolution data.}
    \label{fig:bcd_im}
\end{figure}

The earliest paper with radio continuum observations of Terzan 5 \citep{1990ApJ...365L..63F} reported an integrated flux density of $1.9\pm0.2$ mJy at 1.49 GHz (beam of $18\arcsec \times 10\arcsec$) from C-configuration observations. While this value agrees with their 1.49 GHz A-configuration measurement of diffuse flux, it was not clear to us how far out from the centre this measurement extended, nor whether it successfully separated Ter5C from the diffuse flux. This was our motivation for re-imagining these older data, together with other old data taken in the same setup but a range of configurations, to enable sensitivity to both diffuse flux and the two bright pulsars (see description of this additional imaging in Section \ref{sec:arch}).

In this multi-configuration image (Figure \ref{fig:bcd_im}), which has a large beam ($10.2\arcsec \times 7.3\arcsec$) that is comparable to the core radius, the diffuse central emission is separated from Ter5C\footnote{In this image, Ter5C has a 1.49 GHz flux density of about 1.8 mJy, much more luminous than the value of 1.1 mJy reported by \citet{Martsen2022}, likely due to scintillation.}, and has a flux density consistent with the value of $1.9\pm0.2$ mJy reported by \citet{1990ApJ...365L..63F}. The diffuse emission is not clearly detected beyond the core.

These comparisons of measurements made at different angular resolutions show that there is no evidence our new A-configuration data are missing flux in the core of the cluster, corresponding to the 9.6\arcsec\ radius of our central measurement. We cannot make a confident statement about whether this is also true of our integrated measurement out to 13.3\arcsec, owing to the noise levels in these old data as well as the effects of Ter5C. Since the extrapolated total  measurements from different radii in our data agree to within $\sim 10\%$, to the extent we trust the inner measurement, we can have confidence that the large radius measurement is not too much in error.

We can now proceed to use these unresolved flux measurements in our modelling of the pulsar luminosity function.

\subsection{Analytic Models}

\subsubsection{Power-law Model} \label{sec:pl_mod}

We first fit a hierarchical Bayesian power-law model with free parameter $\alpha_{p}$ (where $dN/dL = L^{-\alpha_{p}}$) to the observed (in the VLA data) individual pulsar flux densities and uncertainties and the observed number of pulsars, assuming the luminosity function is censored below the $3\sigma$ limit of our 2.5 GHz catalogue ($11.2\, \mu$Jy).

These model fits are implemented using the Bayesian Markov Chain Monte Carlo software {\tt JAGS} \citep{2012ascl.soft09002P}. We directly fit the flux densities (rather than converting to pseudo-luminosities), but for ease of reading still refer to a ``luminosity function''. No binning is used.

Fitting the 33 pulsars in our continuum catalogue above this limit, we find $\alpha_{p} = 2.49^{+0.29}_{-0.25}$ if pulsars A and C are excluded, and, as expected, a  flatter slope ($\alpha_{p} = 2.16^{+0.22}_{-0.19}$) if they are included. Note that there is a large gap between Ter5C (319 $\mu$Jy) and the third-brightest pulsar in our data (Ter5P; $51\mu$Jy).

Next we extend the model to also fit the measured unresolved flux below our continuum sensitivity limit. This requires an assumption of the minimum flux density, which we take to be either 0.1 or 1 $\mu$Jy. These values correspond to 2.5 GHz pseudo-luminosities of 3.5 or 35 $\mu$Jy kpc for a distance of 5.9 kpc, equivalent to 1.4 GHz pseudo-luminosities of about 0.01 or 0.1 mJy kpc for an individual pulsar spectral index of $\alpha=1.8$.

We perform model fits within each of the radii defined above (9.6\arcsec\ and 13.3\arcsec) for which we also have measurements of the unresolved flux listed in Table \ref{tab:residual_flux}. Since we also censor the observed pulsars at $11.2\,\mu$Jy, the number of pulsars fit is slightly smaller than discussed above within these radii, 19 and 25 respectively. In these fits, we also obtain a normalisation that corresponds to the total pulsar population $N$, including faint unobserved pulsars. 
 
Assuming a minimum flux density of 1 $\mu$Jy, we find a power-law slope $\alpha_{p} = 2.08^{+0.11}_{-0.10}$ ($\alpha_{p} = 2.05^{+0.09}_{-0.09}$) and an all-cluster pulsar population of $N = 489^{+61}_{-57}$ ($N = 408^{+46}_{-47}$) inferred from the inner 50\% (75\%) samples. For a minimum flux density of 0.1 $\mu$Jy, the best-fit slopes are slightly flatter, with $\alpha_{p} = 1.87\pm0.07$
($\alpha_{p} = 1.85\pm0.07$), but as expected the total pulsar population is much larger, $N = 2091^{+387}_{-367}$ ($N = 1676^{+302}_{-273}$). 

The higher of the two adopted minimum flux densities is likely unrealistic. The faintest pulsar confirmed in Terzan 5 is either Ter5al or Ter5ax, and both have L band flux densities of $\sim 8 \mu$Jy \citep{Cadelano2018,Padmanabh2024}, equivalent to $
\lesssim4\mu$Jy at 2.5 GHz for a typical spectral index.
Given that these pulsars are around the inferred detection limit of current searches \citep{Padmanabh2024}, it is unlikely that they are close to the true minimum flux density. Much less luminous millisecond pulsars are known at nearer distances: for example, the field millisecond pulsar PSR J0318+0253 \citep{Wang2021} would have a 2.5 GHz flux density of $\sim 0.1\,\mu$Jy at the distance of Terzan 5. Adopting this more realistic minimum flux density would imply $\sim 1500$--2000 pulsars, approaching the total maximum population of neutron stars that might plausibly be present in a cluster like Terzan 5 \citep{Ye20}, most of which are unlikely to have been recycled as observable pulsars.

Hence, while we cannot formally reject a power-law model on the basis of available data, it seems much more likely that the pulsars fainter than those found in current searches have a more complex luminosity function.

\subsubsection{Lognormal Model} \label{sec:lognorm_mod}

An alternative model for the luminosity function of pulsars is a lognormal distribution. A low-$N$, censored lognormal distribution can look similar to a power-law, so they are difficult to distinguish when only part of the distribution is observed.

\begin{table}
    \centering
    \renewcommand{\arraystretch}{1.5}
    \caption{Results of lognormal fits to the VLA data. Each radial range has two rows, one for each minimum pulsar flux density assumed (1 or 0.1 $\mu$Jy). The listed uncertainties on the lognormal parameters $\mu$ and $\sigma$ and the total number of pulsars $N$ are equal-tailed 68\% intervals. For $\mu$, because there is support to the low edge of its prior (--2.5), the listed results are prior-dependent: see main text.}    
    \resizebox{\linewidth}{!}{\begin{tabular}{lrccr}\hline\hline
Radius & \multicolumn{1}{c}{Min.} & $\mu$ & $\sigma$ & \multicolumn{1}{c}{$N$} \\
& ($\mu$Jy)  &  (log $\mu$Jy)   & (log $\mu$Jy)  &  \\ \hline
$< 9.6\arcsec$   & 1   & $-1.24^{+0.99}_{-0.85}$ & $0.97^{+0.17}_{-0.24}$ & $419^{+61}_{-53}$ \\
                 & 0.1 & $-1.03^{+0.90}_{-0.96}$ & $0.98^{+0.22}_{-0.26}$ & $1028^{+338}_{-357}$ \\
$< 13.3\arcsec$  & 1   & $-1.36^{+0.91}_{-0.79}$ & $1.02^{+0.16}_{-0.21}$ & $349^{+48}_{-40}$\\  
                & 0.1 & $-1.16^{+0.85}_{-0.87}$ & $1.04^{+0.20}_{-0.23}$ & $875^{+246}_{-256}$ \\\hline
    \end{tabular}}
\begin{flushleft}
\end{flushleft}
    \label{tab:lognorm_fits}
\end{table}

\begin{figure*}
    \centering
    \includegraphics[width=0.99\textwidth]{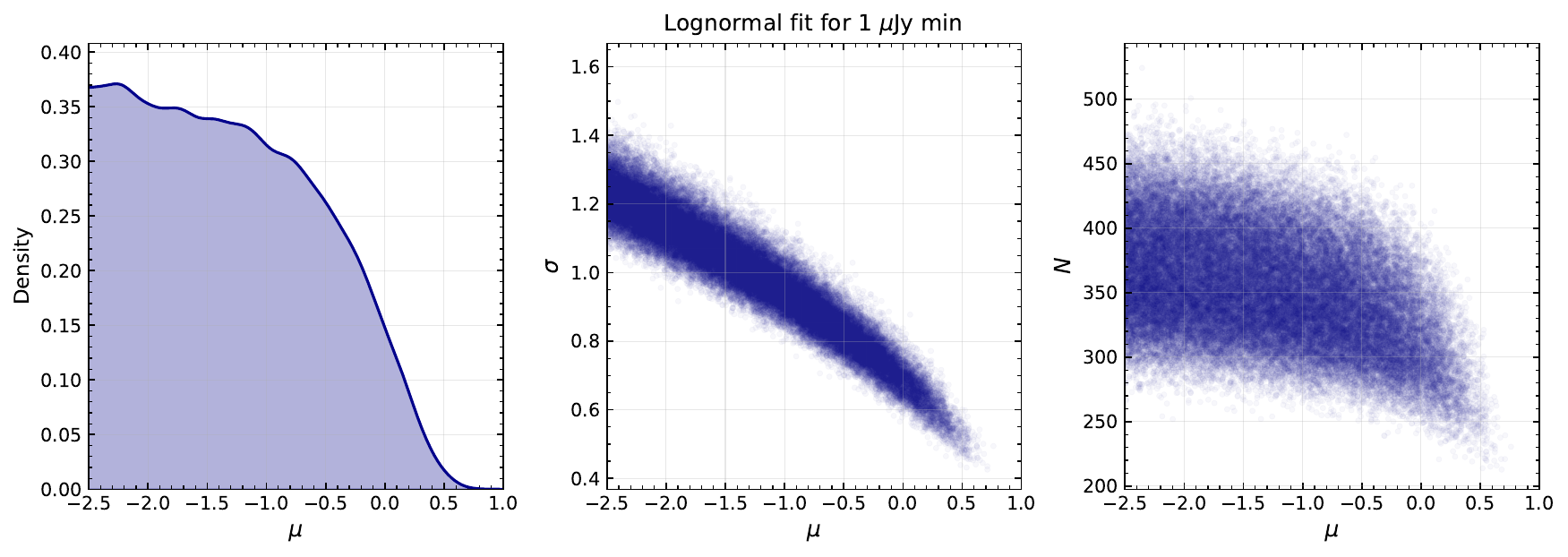}
    \includegraphics[width=0.99\textwidth]{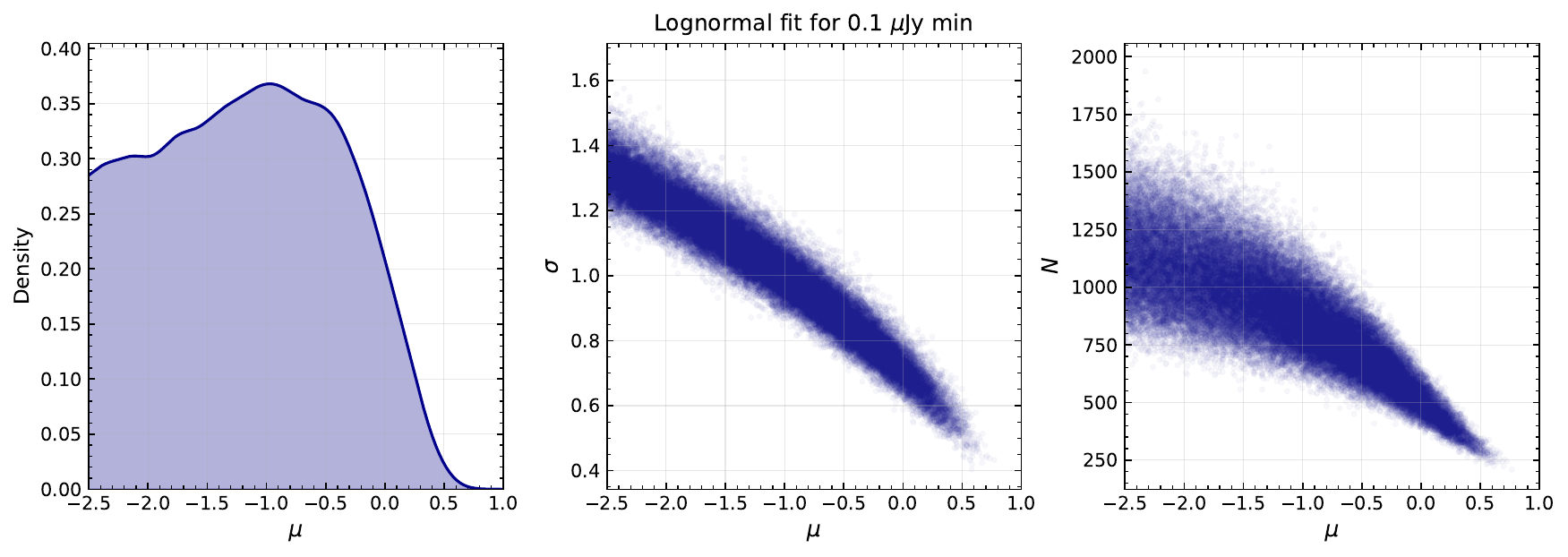}
    \caption{Samples from lognormal fits to the individual pulsar luminosities and the total unresolved flux within a radius of 13.3\arcsec\ (containing $\sim 75$\% of the known pulsars), alternately assuming a minimum flux density of $1 \mu$Jy (top row) or $0.1 \mu$Jy (bottom row.) The left panels are a density estimate of the posterior of the pulsar flux density mean $\mu$. The middle panels show the pulsar flux density standard deviation $\sigma$ versus $\mu$, while the right panels show the total pulsar population $N$ versus $\mu$.}
    \label{fig:log_n}
\end{figure*}

As for the power-law model, we first fit a lognormal model to the individual pulsars without consideration of the unresolved flux. In a lognormal model the log of the flux densities follow a normal distribution with mean $\mu$ and standard deviation $\sigma$. Our fits 
use minimally informative priors on $\mu$ ([--2.5,2.5], corresponding to mean flux densities from about 0.003 $\mu$Jy to 316 $\mu$Jy) and $\sigma$ ([0,4]). These priors are broader than used in other recent papers modelling Terzan 5 pulsars (e.g., \citealt{Chennamangalam2013,Berteaud2024}). Nonetheless,
$\mu$ and $\sigma$ strongly covary and there is still support for $\mu$ to the low end of the prior, even though the prior range is large.  However, we note that high values of $\mu$ that might superficially appear to fit the luminosity function in Figure \ref{fig:lum_fun} (e.g., $\mu \sim 1.3$, $\sigma \sim 0.5$) turn over too quickly at the faint end: they produce only $\sim 100\mu$Jy of unresolved flux, about a factor of 5--7 lower than observed. Hence if the lognormal model is correct, the mean is at lower values.

We next expand the models to include the unresolved flux, again assuming minimum flux densities of either 1 or 0.1 $\mu$Jy and fitting the unresolved flux within the radius containing 50\% or 75\% of the pulsars. The parameters obtained from the two different radii are consistent within the uncertainties. The results of these fits are given in Table \ref{tab:lognorm_fits}, and the results for the larger radius plotted in Figure \ref{fig:log_n}; the plots for the smaller radius are similar.

In all cases, because there is at least some support to the low edge of the $\mu$ prior, the posteriors for $\mu$ and $\sigma$ are  prior-dependent. For this reason, the exact values of $\mu$ and $\sigma$ listed in Table \ref{tab:lognorm_fits} should not be taken too seriously. For all the models there is posterior support for values of $\mu$ that are less than the minimum flux density; in these cases the pulsars would solely be drawn from the right tail of the distribution.

The inferred number of pulsars is less prior-dependent than $\mu$ or $\sigma$, though it does depend on the assumed minimum flux density. In all cases the number of inferred pulsars is large: the lowest median value for any of the fits (75\% radius, $1\, \mu$Jy minimum) is $N = 349^{+48}_{-40}$, while the median $N$ for the more realistic minimum flux density of $0.1 \mu$Jy is at least double this. Even the 1\% quantile of the lowest-$N$ fit is $N=264$, implying that $\gtrsim 200$ pulsars remain to be discovered in Terzan 5. This is a straightforward consequence of the large amount of still-unresolved flux: since it is primarily produced by pulsars fainter than the current detection limit, the number of pulsars needed is large.

In addition to fitting these models with a single censored flux density limit, we also considered more realistic models where the detection probability (in our data) is given by a logistic distribution. This distribution monotonically increases from a low probability of detecting a very faint source to a high probability of detecting a bright pulsar through a transition region of intermediate detection probability. While the detection probability is not known a priori as a function of flux density, we considered a reasonable range of parameters that seemed broadly consistent with our data. For both lognormal and power law models, we found pulsar population numbers slightly lower than but consistent with the values obtained from the simpler models. Since a more complex censoring model has poorly constrained parameters but gives qualitatively similar results, we keep the results from the simpler censoring model fits as those to report. 

\subsection{Image Simulations}

In the previous subsection we found that a lognormal model was consistent with the available data, but that the model parameters were relatively poorly constrained.
Here we consider whether the pixel distribution of the residual flux in the present data contains additional information about the luminosity function of subthreshold sources.

It is definitely the case that the properties of subthreshold pulsars will affect the spatial variations of the residual flux. While a given flux could in principle be produced by a smaller number of brighter pulsars or a larger number of fainter pulsars, these two cases will show differing Poissonian fluctuations, with a larger signal for the smaller number of brighter pulsars; higher angular resolution will also increase the signal, in a manner akin to surface brightness fluctuations in galaxy light \citep{Blakeslee1999}.

In practice, the shot noise signal will be strongly affected by the thermal noise fluctuations in the image, which are larger than the flux densities of most of the faint pulsars in which we are interested. Hence we proceed by a more elaborate route: simulating first the intrinsic pulsar population and then a radio image of this population, which we finally analyse in a manner akin to the original data as described in Section \ref{sec:analysis}.

\subsubsection{Spatial Distribution} \label{sec:spatial_dist}

We have discussed the radii containing an observed fraction of about 50\% or 75\% of the known pulsars, but have not otherwise modelled their radial distribution. To do this, we assume the generalised King model previously used for the distribution of X-ray sources in Terzan 5 \citep{2006ApJ...651.1098H}.
This model assumes a surface density $\Sigma(r) \propto [1+(r/r_c)^2]^{(1-3q)/2}$, where $r$ is the projected radius, $r_c$ is the core radius (for Terzan 5, 9.6\arcsec, identical to observed radius containing half the pulsars in our data), and $q$ is a free parameter that represents the mean mass ratio between the pulsars and the visible cluster stars. Using the 48 millisecond pulsars with precise positions, we find a best-fit value of $q=1.74\pm0.16$. Using this model, the projected radius that contains 50\% (75\%) of the millisecond pulsars is 8.3\arcsec\ (13.3\arcsec), values tolerably close to those observed.

\subsubsection{Simulation methodology}

Pulsars are expected to be point sources superposed on the thermal noise. Thus we first simulate a noise map with rms equal to the observed 2.5\,GHz rms ($3\,\mu\jansky$ beam$^{-1}$) at the location of Terzan\,5. We use the CASA task \verb|simobserve| to simulate our underlying noise map; we use a single noise map for all simulations.

All pulsars are point sources and thus are represented with 2-D Gaussians matching the synthesised beam ($1.4^{\prime\prime}\times0.8^{\prime\prime}$). As in Section \ref{sec:lognorm_mod}, we assume a lognormal analytic model to assign fluxes, but do so over a uniform grid in $\mu$ and $\sigma$: we do not use any of the results from the previous analysis in Section \ref{sec:lognorm_mod} for these simulations. We perform 100 simulations per set of parameters, and assume a minimum flux of $0.1\mu$Jy in all cases.

Source positions are drawn from the previously discussed King model (Section \ref{sec:spatial_dist}). While we experimented with allowing the value of the relative concentration parameter $q$ to vary, we found that this did not significantly affect any results for reasonable variations in $q$. This is perhaps expected given our focus on the core. Hence we fixed $q$ to the previously derived best-fit value of $q=1.74$ for all the results reported below.

To construct the simulated images, simulated sources are added until the cumulative source flux of the cluster within the core ($r<9.6^{\prime\prime}$) matches the observed total (residual plus source) flux within the same radius. This essentially sets the normalisation of our analytic models. Next, all significant ($>3\sigma$) sources are extracted, again in line with our procedure with the real data. We emphasise that to make this process realistic, no prior information on the added sources is used. Hence, due to source blending, the sources extracted will not necessarily match the sources added. After all the significant sources are removed, we have a simulated residual image, whose properties can be compared to the observed residual image. We repeat this process to produce 100 simulated images for each set of parameters.

\subsubsection{Simulation results}
To compare the simulated and real images, we calculate the one-dimensional Kolmorgorov-Smirnov (KS) statistic on the pixel fluxes: this is the maximum difference of the cumulative luminosity functions of the pixel values of two images. Since our focus is on reproducing the observed residual emission, we only consider the pixel flux distributions within the core radius ($\leq9.6^{\prime\prime}$), though the radial distribution of generated sources extends to larger radii in our simulations. Since there are 100 simulated images per parameter pair, there are 100 corresponding KS statistics per parameter pair, as each simulated image is compared to the real image separately. 

\begin{figure}
    \centering
    \includegraphics[width=0.49\textwidth]{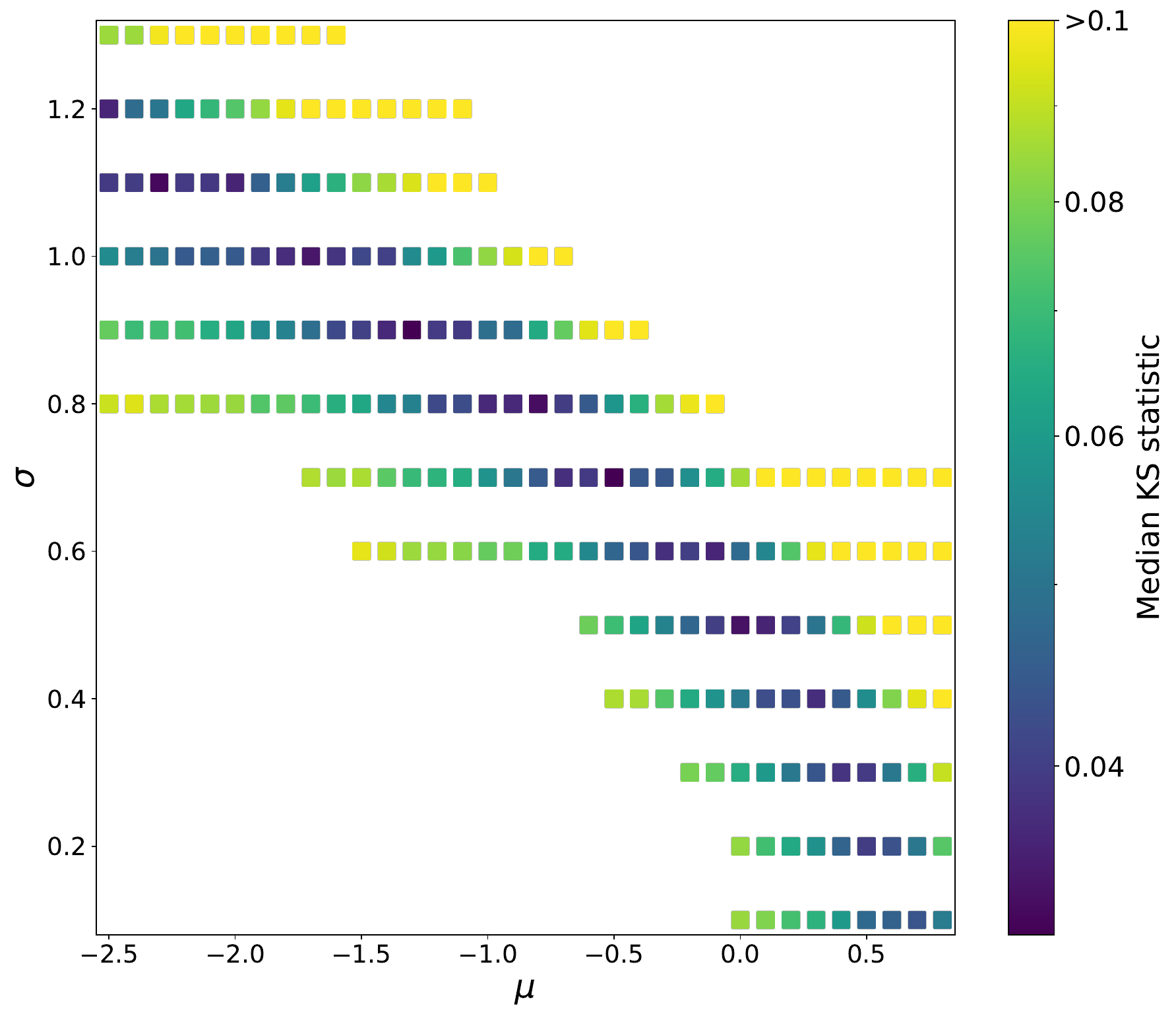}
    \caption{Median KS statistics from comparisons of the simulated and observed residual flux for a range of lognormal parameters, calculated as described in the text. The dark blue/purple simulations are those where the simulated residual images have the most similar pixel flux distribution to the real data.}
    \label{fig:ks}
\end{figure}

In Figure \ref{fig:ks} we show the median KS statistic from comparing the simulated to real data as a function of $\mu$ and $\sigma$. The lowest median KS statistics are found in a broad swath of lognormal model parameter values that appear broadly consistent with the results from the resolved + unresolved luminosity function fitting in Section \ref{sec:lognorm_mod}. In detail, these new image simulations prefer slightly lower $\sigma$ at fixed $\mu$ (or, alternatively, lower $\mu$ at fixed $\sigma$) than the direct luminosity function fitting, but the overall conclusions are similar: a large number of undiscovered pulsars are needed to explain the residual flux. 

We have plotted the KS statistics and not the results from the the KS \emph{test}, and hence the values in Figure \ref{fig:ks} are relative rather than absolute. This is because there are multiple pixels across a synthesised beam, making the individual simulated pixels correlated. Thus a standard one-dimensional KS test (where the samples are independent) is not applicable: that is, our calculated KS statistics cannot be immediately used to accept or reject a null hypothesis that the two pixel distributions are drawn from the same underlying distribution at a particular confidence level.

Instead, to help interpret these KS statistics, we also calculated the KS statistics in a pairwise fashion among the 100 simulations themselves for each $\mu$/$\sigma$  set, making these measurements between the 4950 $\binom{100}{2}$ unique pairs. The typical median KS statistics for these parameter pairs were $\sim 0.03$--0.04. Since these simulations represent identical model parameters, the implication is that when comparing the real image to the simulated images, median KS statistics $\lesssim 0.03-0.04$ imply the observed and simulated residual flux distributions are not inconsistent. This condition is met by the central trough of dark colours in Figure \ref{fig:ks}.

\begin{figure*}
    \centering
    \includegraphics[width=0.495\textwidth]{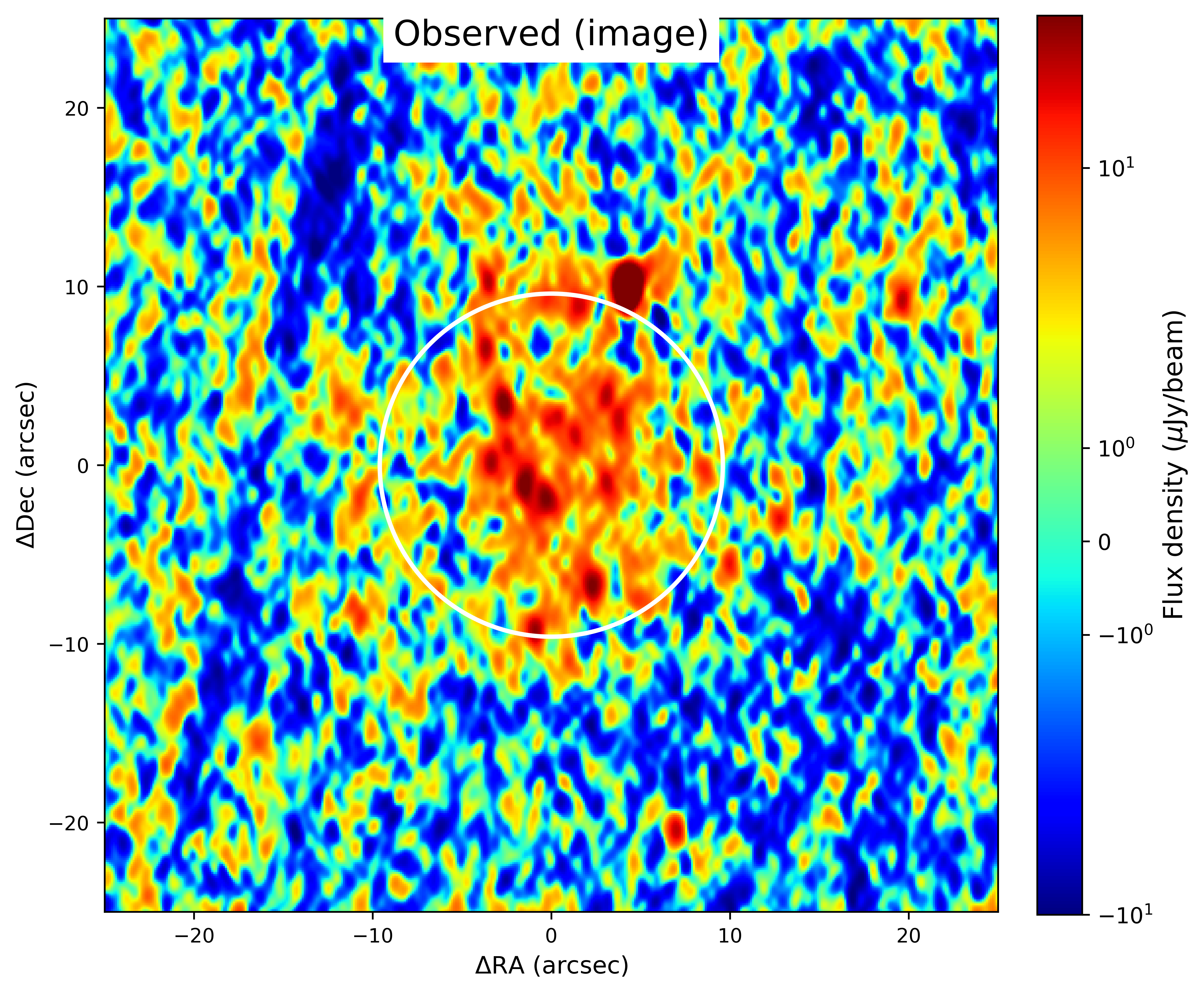}
    \includegraphics[width=0.495\textwidth]{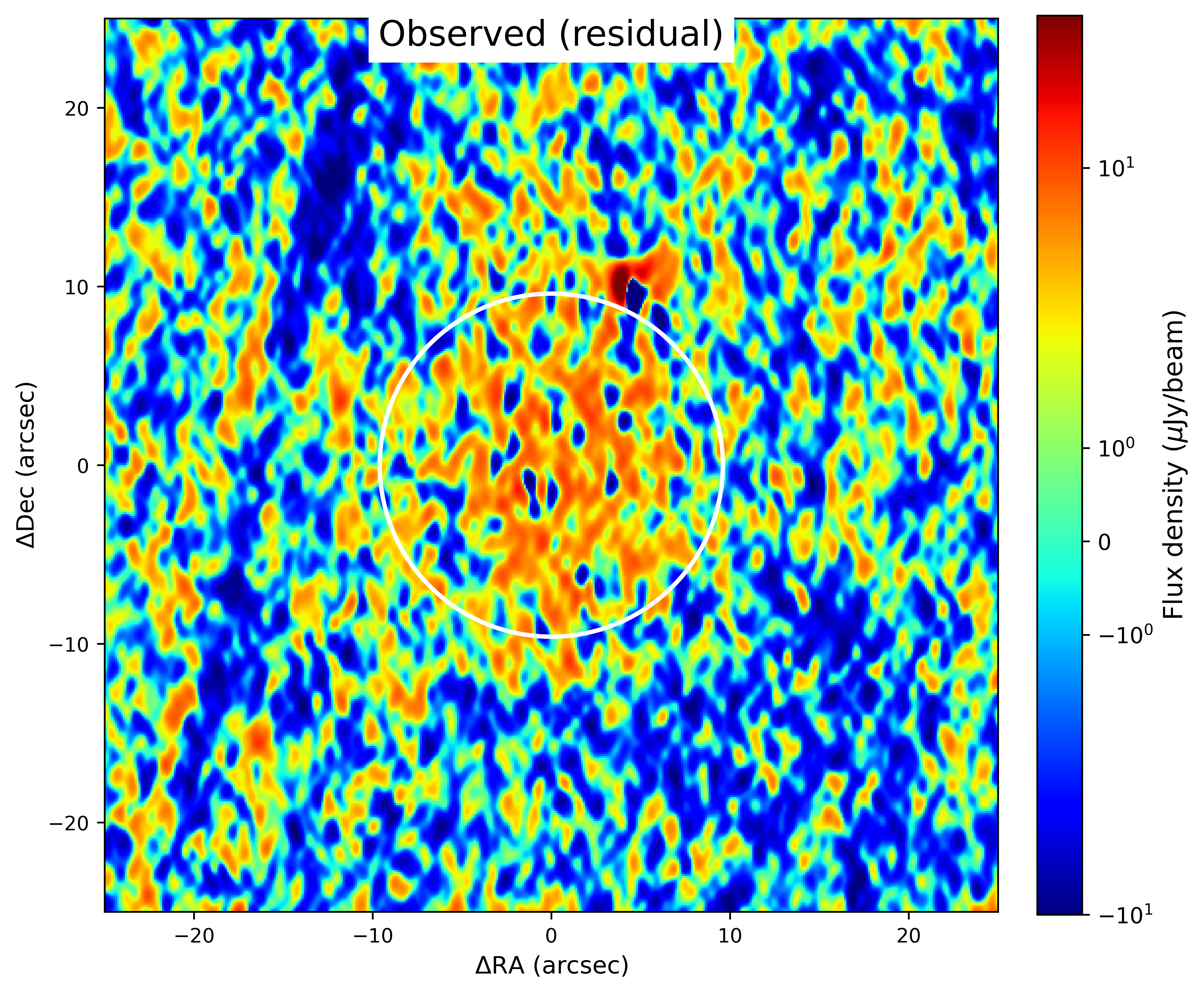}\\
    \includegraphics[width=0.495\textwidth]{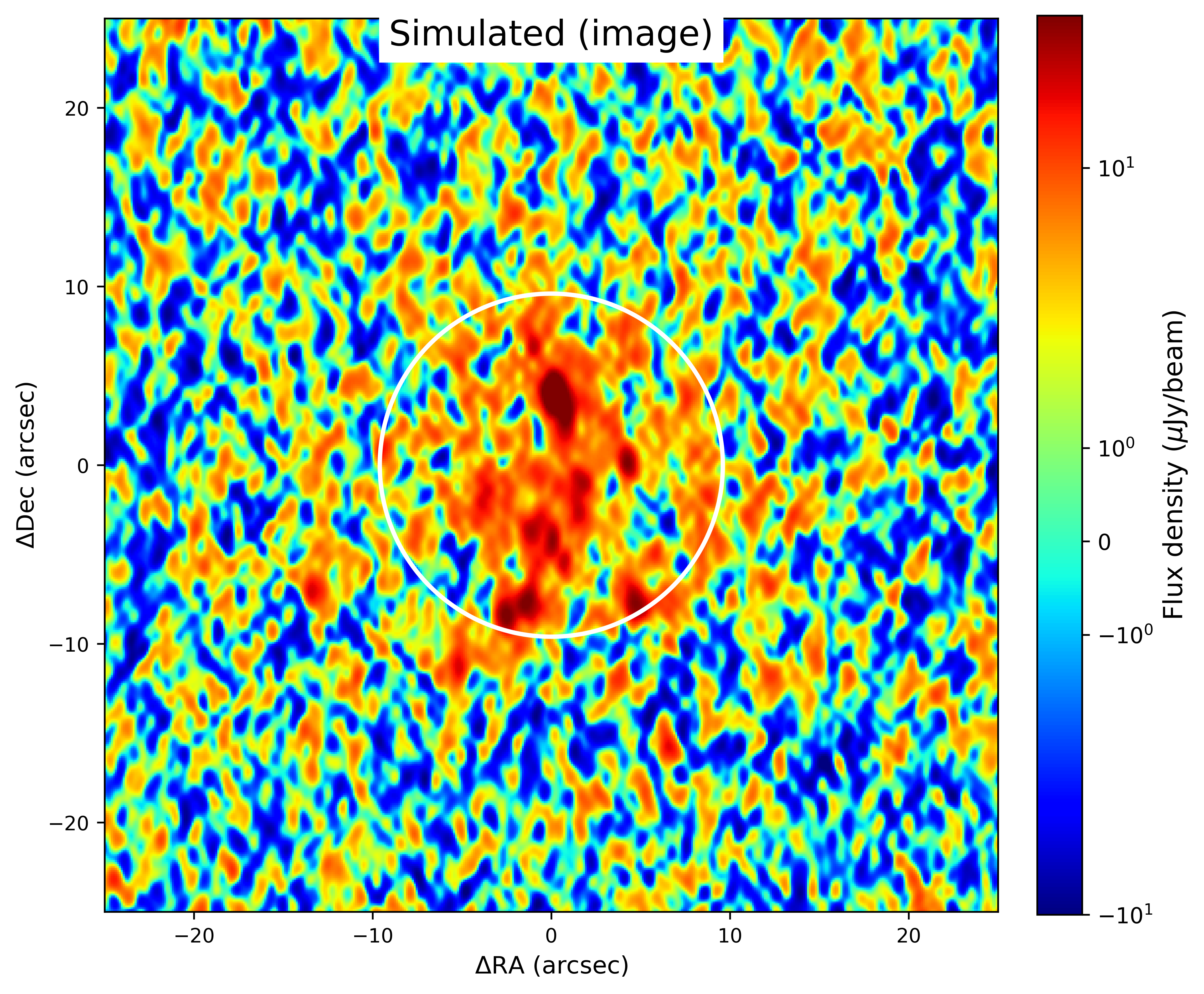}
    \includegraphics[width=0.495\textwidth]{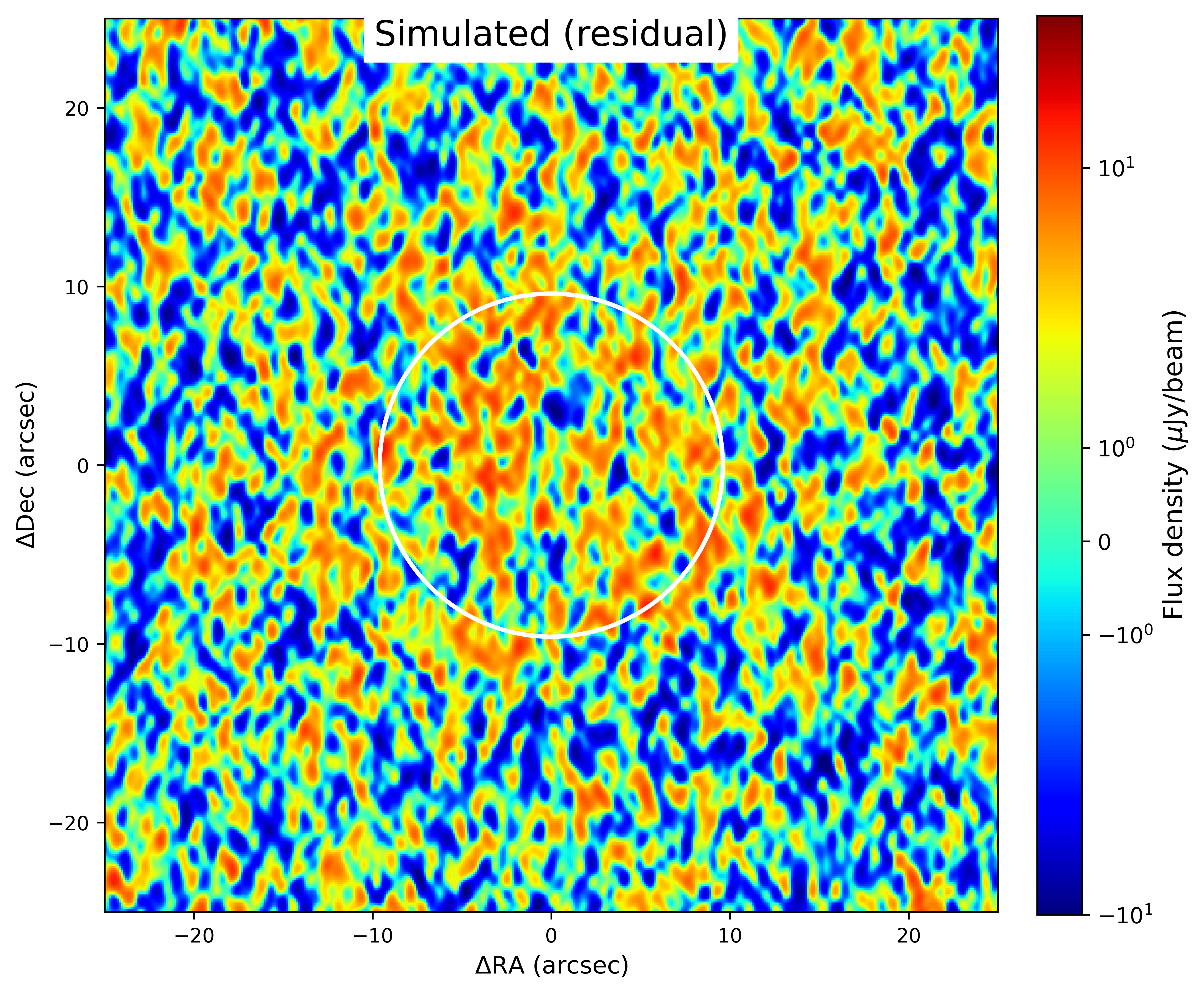}\\
    \includegraphics[width=0.495\textwidth]{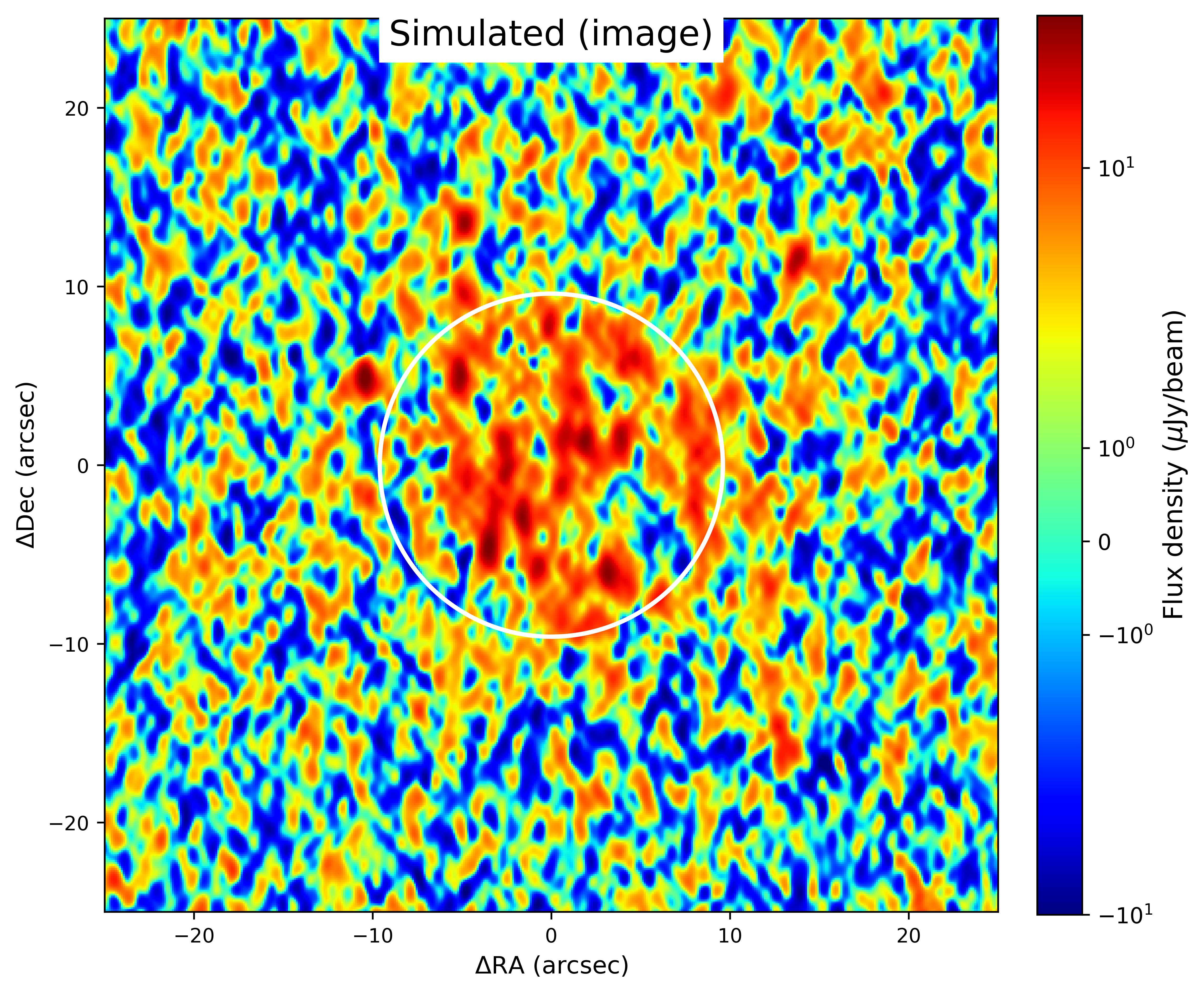}
    \includegraphics[width=0.495\textwidth]{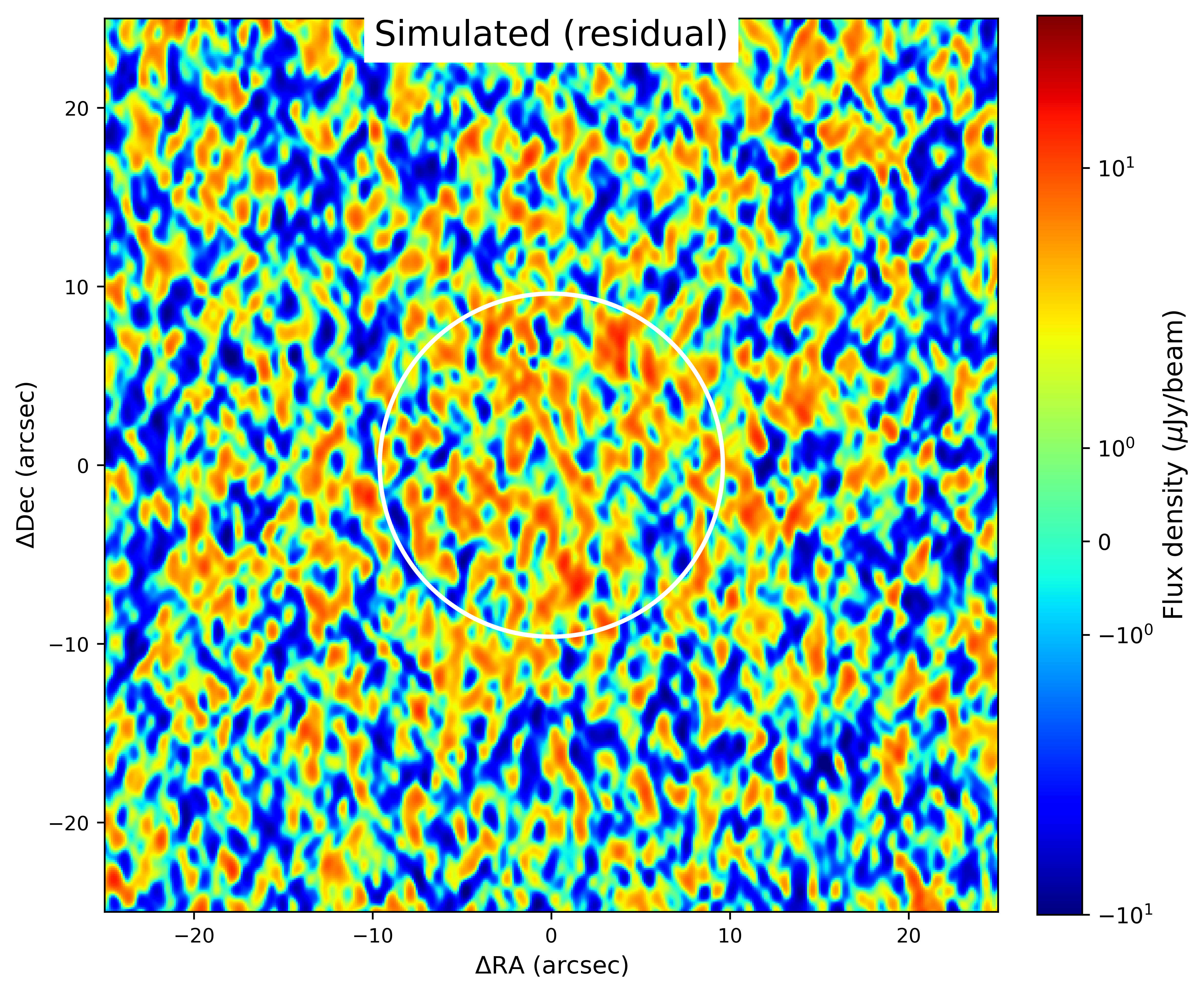}\\
    \caption{Top: real images of Terzan 5 pre (left) and post (right) source subtraction, as already shown in Figure \ref{fig:image} (top) and Figure \ref{fig:residual}. Middle and Bottom: Sample simulated images with residual flux distributions not inconsistent with that observed (top right). The middle row is a parameter pair with a very large total number of pulsars ($N\sim2000$, $\mu=-2.2$, $\sigma=1.1$); the bottom row has a smaller number of pulsars ($N\sim600$, $\mu=0.3$, $\sigma=0.4$). We emphasise that the simulated images do not attempt to reproduce the non-Gaussian noise artifacts seen outside the core in the observed image: the comparison is only in the circled core region.}
    \label{fig:simulations}
\end{figure*}

As examples to illustrate the simulations, in Figure \ref{fig:simulations} we show three pairs of images, with the left side the original or simulated image, and the right side the residual image. The top row is the real data, while the middle and bottom rows represent examples of the range of acceptable simulations. In each case it is clear that the simulated residual images appear to be reasonable matches for the observed residual flux distribution in the core.

\begin{figure}
    \centering
    \includegraphics[width=0.475\textwidth]{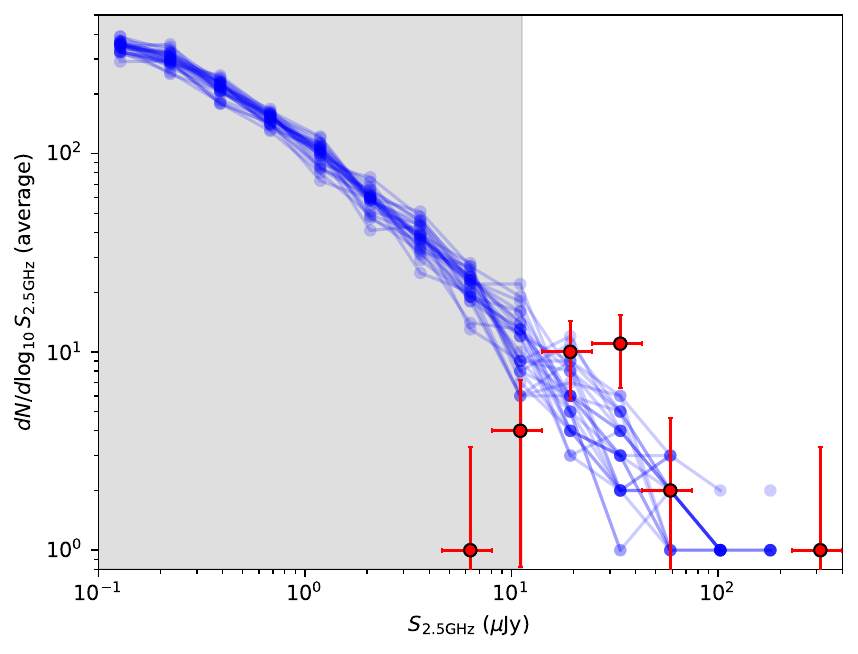}\\
    \includegraphics[width=0.475\textwidth]{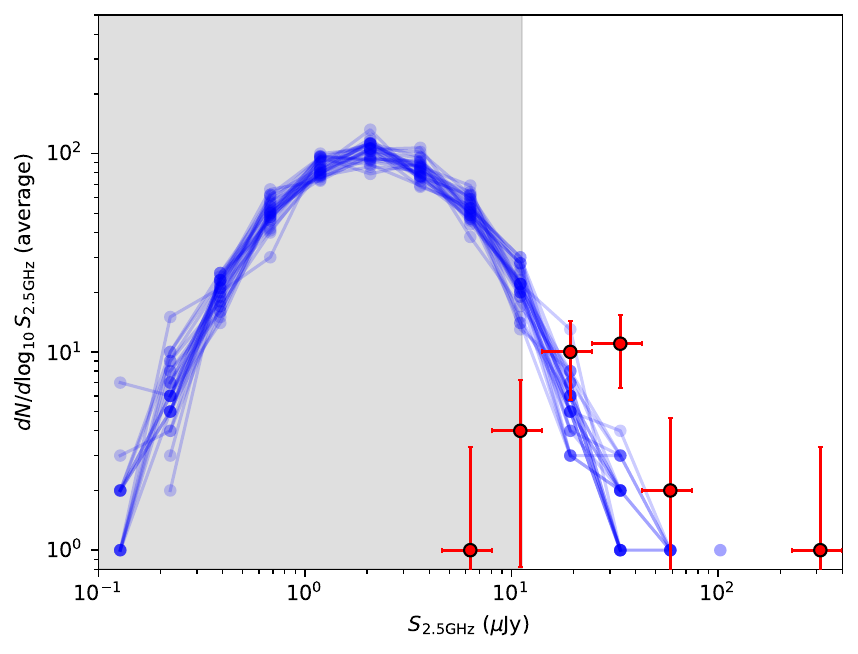}\\    
    \caption{Comparison of observed (red) and simulated (blue) pulsar luminosity functions for the lognormal models shown in Figure \ref{fig:simulations}: the top model has ($\mu=-2.2$, $\sigma=1.1$) while the bottom model has ($\mu=0.3$, $\sigma=0.4$). The observed pulsars are those within $13.3\arcsec$ of the centre. We overlay 25 samples of each model to demonstrate fluctuations, and the grey shaded region indicates the $11.2\mu$Jy incompleteness limit. Both models require a large population of faint undiscovered pulsars, and the extreme larger number of ultra-faint pulsars in the top model is evident.}
    \label{fig:simulated_lumfun}
\end{figure}

Figure \ref{fig:simulated_lumfun} compares the real observed luminosity function to the simulated luminosity function for the same parameter pairs shown in Figure \ref{fig:simulations}. While each model does a reasonable job of representing the brighter detected pulsars, the differing behaviour among the faintest pulsars is why the number of pulsars inferred is so different ($N\sim 2000$ versus $N\sim600$) in the two cases.

Unfortunately, at least here it does not appear that the image simulations offer much stronger constraints on the pulsar luminosity function than modeling the resolved pulsars + unresolved flux: deeper and/or higher resolution data are needed to detect the necessary shot noise signal. We emphasise that these simulations do not \emph{prove} that a lognormal model is the correct one, just that it can produce residual flux distributions not inconsistent with that observed.

\subsection{Gamma-Ray Constraints} 

One disadvantage of using radio data to constrain the pulsar population in Terzan 5 is that there is no straightforward relationship between the radio luminosity and the fundamental physical parameters of the pulsar such as the spindown luminosity. %since the radio emission is energetically irrelevant.

In principle, the $\gamma$-ray emission from Terzan 5 can offer complementary constraints on its pulsars, as the $\gamma$-ray luminosity of a pulsar is essentially uncorrelated with the radio luminosity \citep{Smith2023}. Instead, the $\gamma$-ray luminosity is closely linked to the pulsar spindown luminosity (e.g., \citealt{Abdo2013}). This can be both an advantage and a disadvantage: while $\gamma$-rays give more direct physical information about the pulsars, the integrated $\gamma$-ray flux from a globular cluster can be dominated by even a single source if a young pulsar is present, as in NGC 6624 \citep{Freire2011}. Terzan 5 has no known young pulsars.

\citet{Abdo2010} used the measured $\gamma$-ray luminosity of Terzan 5, an assumed mean spindown luminosity $\dot{E} = 1.8 \times 10^{34}$ erg s$^{-1}$, and $\gamma$-ray conversion efficiency of 8\% to estimate a total population of $N = 180^{+100}_{-90}$ pulsars.

The most updated measurement of the 0.1---100 GeV flux of Terzan 5 is $9.2\pm0.2 \times 10^{-11}$ erg s$^{-1}$ cm$^{-2}$ \citep{Amerio2024}. At a distance of 5.9 kpc, this is equivalent to $L_{\gamma} = 3.8 \times 10^{35}$ erg s$^{-1}$. Notably, this is the largest $\gamma$-ray luminosity of any cluster in the \citet{Amerio2024} sample by $\sim 50$\%, consistent with the fact that Terzan 5 has the largest known population of pulsars of any Galactic cluster. Simply applying the mean $L_{\gamma} = 1.44 \times 10^{33}$ erg s$^{-1}$ assumed in \citet{Abdo2010} to this updated total luminosity gives an estimated $N \sim 260$. 

Using a different methodology, \citet{Amerio2024} find a best-fit mean $L_{\gamma}$ of $1.8 \times 10^{33}$ erg s$^{-1}$ for globular clusters by modelling the $\gamma$-ray emission together with the number of known pulsars and the interaction rates. This would imply $N \sim 180$.
However, given that some clusters modelled (including Terzan 5) likely have many more pulsars than currently known, this reported mean luminosity will be biased higher than the true mean. It is also the case that quite a few field radio millisecond pulsars are not detected as $\gamma$-ray pulsars, suggesting the typical observed $\gamma$-ray efficiency could be higher than the true mean efficiency \citep{Smith2023}. Both factors would tend to reduce the mean $L_{\gamma}$ and lead to a greater inferred pulsar population.

As a limiting case, consider the lognormal fits to the radio luminosities above, which allow a population of $\sim 1000$ pulsars depending on the assumptions. This would require a mean $L_{\gamma}$ of $4 \times 10^{32}$ erg s$^{-1}$ in Terzan 5. While lower than found by \citet{Amerio2024}, this value is close to that inferred from modelling disk millisecond pulsars by \citet{Holst2025}, which is $L_{\gamma} \sim 6 \times 10^{32}$ erg s$^{-1}$.

Another approach is to compare Terzan 5 to a cluster whose millisecond pulsar census is likely more complete: 47 Tuc. This well-studied cluster has a minimum of 42 millisecond pulsars\footnote{https://www3.mpifr-bonn.mpg.de/staff/pfreire/GCpsr.html} and a $\gamma$-ray luminosity that is a factor of $\sim 5.8$ lower than Terzan 5 \citep{Amerio2024}. Making the questionable assumption that the pulsars in the two clusters have the same mean $L_{\gamma}$ would then imply $N \gtrsim 240$ in Terzan 5.

Due to the uncertainties on the $L_{\gamma}$ luminosity function, it does not appear that the $\gamma$-ray properties offer strong independent constraints on the pulsar population in Terzan 5 at present. However, the $\gamma$-ray data do not contradict our radio finding that a large number of pulsars are yet undetected.

\section{Discussion}\label{sec:discussion}

\subsection{Observed Pulsar Luminosity Function}

Based on flux densities measured from pulsar timing over many epochs, \citet{Martsen2022} found that their
 cumulative 1.4 GHz luminosity function showed an apparent change in slope near the faint end of their luminosity function (at a 1.4 GHz flux density of $\sim 23\mu$Jy), which they attributed to observational incompleteness. The most recent Terzan 5 pulsar timing paper, \citet{Padmanabh2024}, briefly revisits the luminosity function of the timed pulsars, again finding a potential slope change in the cumulative luminosity function at a 1.28 GHz flux density $\sim 25\mu$Jy, which they argue is ``suggesting that we are approaching a minimum luminosity cutoff''.

We disagree: we think \citet{Martsen2022} were correct in attributing a slope change to observational incompleteness. From the radiometer equation, \citet{Padmanabh2024} estimate their detection limit to be in the range of $\sim 11$--$17\mu$Jy at 1.28 GHz, depending on the overlap between synthesised beams (they tiled Terzan 5 with 288 separate beams out to a radius of $3^{\prime}$). Uncertainties in the pulsar duty cycle, scintillation, and scattering at L band imply that the detection efficiency at this listed limit is below 100\%. Indeed, given how close this detection limit is to the apparent slope change in the cumulative luminosity function at $\sim 25\,\mu$Jy, observational incompleteness is a more straightforward explanation, rather than an actual change in the luminosity function.  

We emphasise it is \emph{possible} that the luminosity function is indeed changing its slope around the current pulsar detection limits, but that  the presently available data are too uncertain to clearly establish this. This is primarily due to the modest flux range of existing data and the small numbers of pulsars: excluding pulsars A and C, the remaining sources cover at most 1 dex in flux density to the detection limit. Our continuum 2.5 GHz luminosity function (Figure \ref{fig:lum_fun}) allows a change in the slope of the luminosity function near our 2.5 GHz detection limit, but does not demand it.

What is not uncertain is whether current pulsar searches are near a cutoff in the luminosity function. As we have shown, the amount of residual flux is still so large that many pulsars must be undetected--and also much fainter---than known pulsars to self-consistently explain the observations. This does not conflict with other physical constraints or observations: as mentioned above, there is at least one field millisecond pulsar that would have a 2.5 GHz flux density around $0.1 \mu$Jy at the distance of Terzan 5 \citep{Wang2021}. In addition, since the radio luminosity of a pulsar is an insignificant fraction of its spindown luminosity, there is no particular physical reason to expect current radio observations are reaching such a luminosity cutoff. Instead, we need to look to observations themselves to constrain models.

\subsection{Total Pulsar Population: Modelling the Luminosity Function}

\citet{Bagchi2011} found that the observed luminosity functions of pulsars in globular clusters could be fit by either power-law or lognormal functions and that the total number of pulsars in Terzan 5 strongly depended on the assumed model and its parameters (including the minimum pulsar pseudo-luminosity). 

In a follow-up paper, \citet{Chennamangalam2013} fit a lognormal model to the extrapolated 1.4 GHz flux densities of 25 pulsars and the total integrated flux from \citet{2000ApJ...536..865F}, alternatively assuming broad or narrow priors on the parameters of the lognormal distribution. For their ``broad'' priors there is still support to the lower edge of the prior distribution in $\mu$, which appears to truncate the posterior. In any case, for these assumptions they find a median and 95\% interval of $N = 142^{+310}_{-110}$, only marginally consistent with our findings. 

While we are not entirely sure of the reason for this disagreement, it is at least partially due to their modelling of the total (diffuse + point source flux). They assume a total 1.4 GHz flux of 5.2 mJy, of which they attribute 1.9 mJy (37\%) to Ter5A, and overall that known pulsars sum to 4.2 mJy (81\%) of the total assumed flux. Hence there is little flux that needs to be attributed to fainter, yet-undetected pulsars. The value of 5.2 mJy appears to be an underestimate: as we discussed above in Section \ref{sec:diff}, the diffuse 1.4 GHz flux within $\lesssim 10\arcsec$ from \citet{2000ApJ...536..865F} is $\sim2$ mJy. The comparison to the lower resolution imaging of \citet{1990ApJ...365L..63F}, and our reimaging of this lower resolution data (Section \ref{sec:diff}), both show that these diffuse flux measurements only pertain to the cluster core. As this radius contains only about half the known pulsars, the diffuse flux integrated over all radii should be double this value, $\sim 4$ mJy. Adding in pulsars Ter5A and Ter5C, visible as point sources in the 1.4 GHz image, gives a total flux of 6.7--7.8 mJy depending on whether the flux densities from this image or \citet{Martsen2022} are assumed for the two bright pulsars. This would imply a much larger amount of ``residual" diffuse flux to be attributed to faint pulsars, leading to a larger N. As our independent measurement of the diffuse flux is consistent with that of \citet{2000ApJ...536..865F} 
(see Section \ref{sec:diff}), we believe this larger value of the diffuse flux is the correct one, implying a need for more pulsars than inferred by \citet{Chennamangalam2013}.

\citet{Berteaud2024} compared Bayesian and simulation-based inference for inferring the pulsar population of a globular cluster, using Terzan 5 as a case study. While they used flux densities for a larger sample of pulsars, they assumed the same diffuse flux as in \citet{Chennamangalam2013}, and found broadly similar results, with a total pulsar population of $N=158^{+294}_{-104}$. They emphasised that their results were extremely sensitive to the diffuse flux and that new measurements of this quantity, as well as deeper surveys, were needed to improve constraints on the pulsar population. This is borne out by our results.

The specific number of pulsars inferred from our fits depend on the models assumed. It may be that the luminosity function is more complex than a lognormal; for example, a broken power law would add another free parameter, and of course yet more complex functions could be considered. But these are not yet demanded by the data.

Overall, we conclude our flux measurements of individual pulsars alone have not substantially improved our understanding of the luminosity function of pulsars in Terzan 5: the main advance is the simultaneous measurement of these with the residual diffuse flux. For any of our model fits, the mean flux is fainter than current detection limits, implying that the specific luminosity function parameters are not well-constrained, and the total number of pulsars is still very uncertain. Nevertheless the pulsar population appears much larger than inferred from previous work, with a ``minimum" reasonable estimate of pulsars of $\sim 250$ and a true value that is likely higher.

\section{Conclusions and Future Work} \label{sec:conclusion}

We have presented new deep high-resolution S-band (2--4 GHz) imaging of Terzan 5, detecting 38 of its 49 timed pulsars as continuum sources. Our central result is that even after subtracting these pulsars from the images, there is still a substantial diffuse residual flux associated with a large population of pulsars below our detection limit. Our data alone are consistent with a pulsar population from $\sim 250$ up to $\gtrsim 1000$ pulsars, though the values at the high end may be unlikely or impossible from a pulsar formation standpoint.

While with the present data the image simulations did not lead to meaningfully stronger conclusions about the pulsar population in Terzan 5, that should change with the ultra-deep, long baseline continuum imaging enabled by the next-generation VLA \citep{Murphy2018} and SKA1-Mid \citep{Swart2022}. We will explore simulated ngVLA observations of Terzan 5 in a future work, also adding physical effects such as scintillation not yet included in our simulations.

However, we need not wait for next generation facilities to make progress: the continued detection of timed pulsars at the sensitivity limits of existing data \citep{Padmanabh2024} already indicates the utility of ongoing searches for new pulsars in Terzan 5. Many pulsars just slightly fainter than those known are ripe for discovery.

\section*{Acknowledgements}

We acknowledge the comments of an anonymous referee, which helped improve the paper. We acknowledge support from NASA grant 80NSSC21K0628, NSF grants AST-2107070, AST-2205550, and AST-2205631 and the Packard Foundation. The National Radio Astronomy Observatory and Green Bank Observatory are facilities of the U.S. National Science Foundation operated under cooperative agreement by Associated Universities, Inc. SMR is a CIFAR Fellow and is supported by the NSF Physics Frontiers Center award 2020265 and NSF AAG award 2510064. COH is supported by NSERC Discovery Grant RGPIN-2023-04264 and Alberta Innovates Advance Program \#242506334.

\section*{Data Availability}

The VLA data are available via the NRAO data archive (Project code 22A-396) \url{https://data.nrao.edu/portal/}.



\bibliographystyle{mnras}
\bibliography{references} % if your bibtex file is called example.bib

%\begin{thebibliography}{99}
%\end{thebibliography}

%%%%%%%%%%%%%%%%%%%%%%%%%%%%%%%%%%%%%%%%%%%%%%%%%%

%%%%%%%%%%%%%%%%% APPENDICES %%%%%%%%%%%%%%%%%%%%%

\appendix

\setcounter{table}{0}
\renewcommand{\thetable}{A\arabic{table}}
\captionsetup{width=1\textwidth}

\onecolumn
\renewcommand{\arraystretch}{1.5} % Increase row spacing by 50%
\begin{longtable}{lccccccccccc}
\caption{Catalog of Terzan\,5 sources. The columns are the (1) source ID number, (2--5) position and positional uncertainty, (6) alternative name for previously known sources (Ter5 CX1 is the transitional millisecond pulsar candidate \citep{2018ApJ...864...28B} and EXO 1745-248 is a transient low-mass X-ray binary \citep{2016MNRAS.460..345T}), (7) distance from the cluster core, (8--9) source flux densities or 3$\sigma$ upper limit for non-detection, (10) spectral index, (11) 2012 2.6\,GHz flux density from U20, (12) 1--10 keV X-ray luminosity from \citet{goose}.}\\
\hline\hline
\# & RA (ICRS) & $\sigma_{\mathrm RA}$ & Dec (ICRS) & $\sigma_{\mathrm Dec}$ &Alt ID &  Rad. & $S_{2.5\,GHz}$ & $S_{3.5\,GHz}$ &  $\alpha$ & $S_{2.6\,GHz}$ &$L_{\mathrm X}$\\
 & h:m:s & \arcsec & $^{\circ}$:$^{\prime}$:$^{\prime\prime}$ & \arcsec &  &  \arcmin & \ujy &\ujy & & \ujy & $10^{31}$\ergs \\\hline
\endfirsthead

\multicolumn{5}{l}
{ \tablename\ \thetable{} -- continued from previous page} \\\hline\hline
\# & RA (ICRS) & $\sigma_{\mathrm RA}$ & Dec (ICRS) & $\sigma_{\mathrm Dec}$ &Alt ID &  Rad. & $S_{2.5\,GHz}$ & $S_{3.5\,GHz}$ &  $\alpha$ & $S_{2.6\,GHz}$ &$L_{\mathrm X}$\\
 & h:m:s & \arcsec & $^{\circ}$:$^{\prime}$:$^{\prime\prime}$ & \arcsec &  &  \arcmin & \ujy &\ujy & & \ujy & $10^{31}$\ergs \\\hline
\endhead

\hline \multicolumn{5}{l}{Continued on next page}
\endfoot

\hline
\endlastfoot

1&17:48:02.247&0.06&-24:46:37.82&0.11&Ter5A&0.60&714$\pm$3&234$\pm$2&$-3.31_{-0.03}^{+0.03}$&698$\pm$7&$0.05_{-0.05}^{+1.29}$\\
2&17:47:50.917&0.06&-24:44:51.39&0.11&--&3.68&363$\pm$3&299$\pm$2&$-0.58_{-0.03}^{+0.03}$&327$\pm$7&--\\
3&17:48:06.404&0.06&-24:45:04.95&0.11&--&1.70&243$\pm$3&200$\pm$3&$-0.59_{-0.06}^{+0.05}$&287$\pm$6&--\\
4&17:48:10.921&0.06&-24:45:40.13&0.11&--&1.75&146$\pm$3&145$\pm$3&$-0.03_{-0.09}^{+0.09}$&192$\pm$6&--\\
5&17:48:05.041&0.06&-24:46:41.30&0.11&Ter5P&0.07&51$\pm$4&29$\pm$3&$-1.7_{-0.3}^{+0.3}$&224$\pm$7&$43.8_{-1.6}^{+1.7}$\\
6&17:47:55.637&0.06&-24:44:58.97&0.11&--&2.73&126$\pm$3&95$\pm$2&$-0.8_{-0.1}^{+0.1}$&173$\pm$6&--\\
7&17:48:04.534&0.06&-24:46:34.83&0.11&Ter5C&0.18&319$\pm$4&116$\pm$3&$-3.01_{-0.08}^{+0.07}$&275$\pm$7&--\\
8&17:48:09.899&0.06&-24:48:57.66&0.11&--&2.50&94$\pm$3&87$\pm$3&$-0.2_{-0.1}^{+0.1}$&108$\pm$6&--\\
9&17:48:05.106&0.06&-24:46:34.51&0.11&Ter5V&0.18&29$\pm$4&21$\pm$3&$-1.0_{-0.5}^{+0.6}$&93$\pm$7&$2.6_{-1.6}^{+3.1}$\\
10&17:48:16.813&0.06&-24:46:41.59&0.11&--&2.72&82$\pm$3&71$\pm$4&$-0.4_{-0.2}^{+0.2}$&60$\pm$6&--\\
11&17:47:53.154&0.06&-24:46:52.01&0.11&--&2.66&67$\pm$3&57$\pm$2&$-0.5_{-0.2}^{+0.2}$&71$\pm$6&--\\
12&17:47:57.125&0.06&-24:45:21.08&0.11&--&2.24&67$\pm$3&42$\pm$2&$-1.4_{-0.2}^{+0.2}$&76$\pm$6&--\\
13&17:48:08.874&0.06&-24:47:34.14&0.11&--&1.23&40$\pm$3&32$\pm$3&$-0.6_{-0.3}^{+0.3}$&54$\pm$6&--\\
14&17:48:01.734&0.06&-24:46:28.13&0.11&--&0.76&58$\pm$3&53$\pm$2&$-0.2_{-0.2}^{+0.2}$&41$\pm$7&--\\
15&17:48:08.064&0.06&-24:46:00.98&0.11&--&1.03&43$\pm$3&31$\pm$2&$-1.0_{-0.3}^{+0.3}$&21$\pm$7&--\\
16&17:47:49.281&0.06&-24:46:19.72&0.11&--&3.56&44$\pm$3&37$\pm$2&$-0.5_{-0.3}^{+0.3}$&38$\pm$6&--\\
17&17:48:04.623&0.06&-24:46:40.80&0.11&Ter5M&0.08&31$\pm$4&10$\pm$3&$-3.0_{-0.4}^{+0.5}$&58$\pm$7&$19.7_{-2.7}^{+2.9}$\\
18&17:48:04.956&0.06&-24:46:45.66&0.11&Ter5ae/Z&0.03&56$\pm$4&24$\pm$3&$-2.5_{-0.4}^{+0.4}$&66$\pm$7&$1.9_{-1.0}^{+2.7}$\\
19&17:47:52.460&0.06&-24:48:58.42&0.11&--&3.59&36$\pm$3&28$\pm$2&$-0.7_{-0.4}^{+0.4}$&38$\pm$7&--\\
20&17:47:50.438&0.06&-24:46:34.73&0.11&--&3.28&45$\pm$3&39$\pm$2&$-0.4_{-0.2}^{+0.3}$&39$\pm$6&--\\
21&17:48:17.922&0.16&-24:46:31.894&0.14&--&2.97&$<$11&$<$11&--&26$\pm$7&--\\
22&17:48:05.100&0.06&-24:46:44.40&0.11&Ter5Y&0.06&35$\pm$4&13$\pm$3&$-2.7_{-0.5}^{+0.6}$&46$\pm$7&$2.3_{-0.4}^{+0.5}$\\
23&17:48:08.566&0.08&-24:44:15.75&0.14&--&2.62&28$\pm$3&$<$8&--&--&--\\
24&17:48:14.271&0.08&-24:46:50.93&0.14&--&2.14&28$\pm$3&$<$9&--&38$\pm$6&--\\
25&17:48:15.084&0.16&-24:47:47.332&0.14&--&2.55&$<$10&$<$10&--&35$\pm$6&--\\
26&17:48:04.869&0.06&-24:46:46.55&0.11&Ter5I&0.03&45$\pm$4&20$\pm$3&$-2.5_{-0.4}^{+0.4}$&--&--\\
27&17:48:04.915&0.06&-24:46:53.87&0.11&Ter5N&0.16&36$\pm$4&24$\pm$3&$-1.2_{-0.4}^{+0.4}$&62$\pm$7&$0.5_{-0.5}^{+2.1}$\\
28&17:48:03.407&0.06&-24:46:35.57&0.11&Ter5E&0.36&28$\pm$3&19$\pm$2&$-1.0_{-0.5}^{+0.5}$&46$\pm$7&$0.02_{-0.01}^{+0.13}$\\
29&17:48:05.116&0.06&-24:46:38.13&0.11&Ter5F&0.12&31$\pm$4&20$\pm$3&$-1.3_{-0.5}^{+0.5}$&41$\pm$7&$0.02_{-0.02}^{+0.17}$\\
30&17:48:04.677&0.06&-24:46:51.45&0.11&Ter5O&0.12&43$\pm$4&13$\pm$3&$-3.1_{-0.3}^{+0.4}$&53$\pm$7&$2.3_{-0.4}^{+0.4}$\\
31&17:48:05.023&0.06&-24:46:43.52&0.11&--&0.04&27$\pm$4&17$\pm$3&$-1.4_{-0.7}^{+0.6}$&--&$2.1_{-0.5}^{+0.5}$\\
32&17:48:04.734&0.06&-24:46:35.90&0.11&Ter5L&0.15&27$\pm$4&15$\pm$3&$-1.6_{-0.7}^{+0.7}$&30$\pm$7&$0.03_{-0.02}^{+0.19}$\\
33&17:48:03.904&0.06&-24:46:47.69&0.11&Ter5K&0.22&20$\pm$4&9$\pm$3&$-2.2_{-0.8}^{+0.9}$&24$\pm$7&$0.02_{-0.02}^{+0.19}$\\
34&17:48:04.677&0.16&-24:46:48.743&0.14&--&0.08&$<$11&$<$8&--&--& $8.0_{-0.8}^{+0.8}$\\
35&17:48:04.572&0.06&-24:46:42.22&0.11&CX1&0.07&30$\pm$4&13$\pm$3&$-2.5_{-0.6}^{+0.6}$&50$\pm$7&$43.9_{-1.6}^{+1.9}$\\
36&17:48:04.754&0.06&-24:46:43.06&0.11&Ter5ab&0.03&28$\pm$4&10$\pm$3&$-2.7_{-0.5}^{+0.7}$&53$\pm$7&$10.2_{-1.8}^{+1.7}$\\
37&17:48:10.329&0.06&-24:47:59.08&0.11&--&1.76&35$\pm$3&26$\pm$3&$-0.9_{-0.4}^{+0.4}$&43$\pm$6&--\\
38&17:48:04.618&0.06&-24:46:45.71&0.11&Ter5ar&0.06&30$\pm$4&18$\pm$3&$-1.6_{-0.6}^{+0.6}$&38$\pm$7&$25.5_{-1.3}^{+1.3}$\\
39&17:48:05.924&0.06&-24:46:05.71&0.11&Ter5D&0.69&28$\pm$3&14$\pm$2&$-2.0_{-0.6}^{+0.6}$&37$\pm$7&$0.01_{-0.01}^{+0.07}$\\
40&17:48:04.418&0.16&-24:46:48.780&0.14&--&0.12&$<$11&$<$8&--&36$\pm$7&$1.5_{-0.4}^{+0.9}$\\
41&17:48:03.726&0.06&-24:47:23.45&0.11&--&0.70&12$\pm$3&9$\pm$2&$-0.6_{-1.2}^{+1.1}$&31$\pm$7&--\\
42&17:48:05.225&0.08&-24:46:47.661&0.14&EXO1745&0.10&$<$11&$<$8&--&--&$179_{-3}^{+3}$\\
43&17:48:04.835&0.08&-24:46:42.17&0.14&Ter5W&0.04&28$\pm$4&$<$8&--&--&--\\
44&17:48:12.548&0.08&-24:43:55.59&0.14&--&3.31&28$\pm$3&$<$10&--&--&--\\
45&17:47:54.477&0.06&-24:45:45.17&0.11&--&2.55&27$\pm$3&23$\pm$2&$-0.4_{-0.4}^{+0.4}$&--&--\\
46&17:48:04.335&0.06&-24:47:05.14&0.11&Ter5Q&0.36&26$\pm$3&13$\pm$2&$-2.0_{-0.7}^{+0.7}$&--&$0.06_{-0.05}^{+0.16}$\\
47&17:48:08.930&0.08&-24:47:34.84&0.14&--&1.25&25$\pm$3&$<$8&--&--&--\\
48&17:47:59.783&0.06&-24:45:33.31&0.11&--&1.65&25$\pm$3&16$\pm$2&$-1.2_{-0.5}^{+0.5}$&--&--\\
49&17:47:55.430&0.06&-24:47:42.71&0.11&--&2.35&24$\pm$3&25$\pm$2&$0.2_{-0.4}^{+0.5}$&--&--\\
50&17:47:57.263&0.06&-24:49:29.60&0.11&--&3.24&24$\pm$3&19$\pm$2&$-0.6_{-0.6}^{+0.6}$&--&--\\
51&17:47:51.719&0.06&-24:48:26.35&0.11&--&3.43&24$\pm$3&21$\pm$2&$-0.4_{-0.5}^{+0.5}$&--&--\\
52&17:48:15.773&0.08&-24:45:58.91&0.14&--&2.59&23$\pm$3&$<$10&--&--&--\\
53&17:48:04.915&0.06&-24:46:47.35&0.11&Ter5au&0.05&22$\pm$4&10$\pm$3&$-2.3_{-0.7}^{+0.8}$&--&--\\
54&17:47:59.487&0.08&-24:46:23.84&0.14&--&1.27&21$\pm$3&$<$6&--&--&--\\
55&17:47:55.570&0.06&-24:44:17.83&0.11&--&3.23&21$\pm$3&17$\pm$2&$-0.7_{-0.6}^{+0.6}$&--&--\\
56&17:47:55.263&0.08&-24:45:14.34&0.14&--&2.65&21$\pm$3&$<$6&--&--&--\\
57&17:47:53.230&0.06&-24:47:42.16&0.11&--&2.81&20$\pm$3&18$\pm$2&$-0.3_{-0.6}^{+0.6}$&--&--\\
58&17:47:53.147&0.08&-24:46:29.14&0.14&--&2.67&20$\pm$3&$<$6&--&--&--\\
59&17:48:01.620&0.08&-24:44:57.77&0.14&--&1.93&19$\pm$3&$<$7&--&--&--\\
60&17:48:04.114&0.06&-24:46:50.06&0.11&Ter5ai&0.19&19$\pm$4&16$\pm$3&$-0.5_{-0.8}^{+0.8}$&--&--\\
61&17:47:59.040&0.08&-24:47:00.76&0.14&--&1.35&19$\pm$3&$<$7&--&--&--\\
62&17:48:04.862&0.08&-24:46:35.43&0.16&--&0.15&18$\pm$4&$<$8&--&--&--\\
63&17:48:05.641&0.08&-24:46:53.03&0.16&Ter5H&0.23&18$\pm$4&$<$8&--&--&$0.02_{-0.01}^{+0.15}$\\
64&17:48:04.219&0.08&-24:46:44.99&0.17&Ter5af&0.14&17$\pm$4&$<$8&--&--&$0.02_{-0.01}^{+0.14}$\\
65&17:48:04.725&0.08&-24:46:50.86&0.17&Ter5R&0.11&16$\pm$4&$<$8&--&--&--\\
66&17:48:04.775&0.06&-24:46:34.74&0.11&Ter5ag&0.16&16$\pm$4&10$\pm$3&$-1.4_{-1.1}^{+1.1}$&--&--\\
67&17:48:03.791&0.08&-24:47:22.64&0.14&--&0.68&16$\pm$3&$<$7&--&--&--\\
68&17:48:05.650&0.06&-24:46:46.75&0.11&Ter5G&0.18&15$\pm$4&12$\pm$3&$-0.7_{-1.0}^{+1.1}$&--&$0.02_{-0.01}^{+0.10}$\\
69&17:48:05.606&0.08&-24:47:12.02&0.16&Ter5X&0.49&15$\pm$3&$<$7&--&--&$0.02_{-0.01}^{+0.11}$\\
70&17:48:04.489&0.08&-24:46:52.23&0.19&Ter5ax&0.15&15$\pm$4&$<$8&--&--&--\\
71&17:48:07.416&0.08&-24:47:04.46&0.16&--&0.67&15$\pm$3&$<$8&--&--&--\\
72&17:48:05.562&0.08&-24:46:06.75&0.17&--&0.65&14$\pm$3&$<$7&--&--&--\\
73&17:48:04.008&0.08&-24:47:40.66&0.17&Ter5J&0.95&13$\pm$3&$<$7&--&--&--\\
74&17:48:05.656&0.08&-24:46:41.84&0.14&Ter5an&0.19&13$\pm$4&$<$8&--&--&\\
75&17:48:05.149&0.06&-24:46:35.94&0.11&Ter5ao&0.16&11$\pm$4&11$\pm$3&$-0.2_{-1.3}^{+1.2}$&--&--\\
76&17:48:03.017&0.08&-24:46:52.62&0.17&Ter5T&0.44&9$\pm$3&$<$7&--&--&--\\
77&17:48:05.805&0.08&-24:46:42.39&0.14&Ter5aa&0.22&9$\pm$4&$<$8&--&--&$0.02_{-0.01}^{+0.14}$\\
78&17:48:05.031&0.06&-24:46:35.27&0.12&Ter5aj&0.16&9$\pm$4&11$\pm$3&$0.4_{-1.2}^{+1.0}$&--&$0.05_{-0.04}^{+0.31}$\\
79&17:48:06.045&0.08&-24:46:32.97&0.23&Ter5ac&0.33&8$\pm$3&$<$8&--&--&--\\
80&17:48:04.311&0.08&-24:46:41.76&0.22&Ter5ah&0.13&8$\pm$4&$<$8&--&--&--\\
81&17:48:04.295&0.08&-24:46:32.22&0.23&Ter5S&0.24&7$\pm$4&$<$8&--&--&$0.02_{-0.01}^{+0.13}$\\
82&17:48:11.830&0.06&-24:44:09.47&0.11&--&3.03&$<$10&22$\pm$3&--&--&--\\
83&17:48:14.656&0.06&-24:46:03.14&0.11&--&2.33&$<$10&21$\pm$3&--&--&--\\
84&17:48:12.031&0.06&-24:45:49.38&0.11&--&1.87&$<$10&20$\pm$3&--&--&--\\
85&17:47:56.866&0.06&-24:43:33.40&0.11&--&3.67&$<$9&19$\pm$2&--&--&--\\
86&17:48:04.684&0.06&-24:46:35.50&0.12&--&0.16&$<$11&12$\pm$3&--&--&--\\
87&17:48:06.021&0.06&-24:47:23.26&0.13&--&0.70&$<$9&11$\pm$2&--&--&--\\
88&17:48:05.045&0.06&-24:46:44.36&0.14&--&0.04&$<$11&11$\pm$3&--&--&--\\
89&17:48:04.842&0.06&-24:46:45.77&0.17&--&0.02&$<$11&9$\pm$3&--&--&$9.8_{-3.6}^{+7.4}$\\
\label{tab:src_list}
\end{longtable}

\onecolumn
\captionsetup{width=0.56\textwidth}

\begin{longtable}{lccllc}
\caption{Radio light curve of Ter5A plotted in Figure \ref{fig:lc_ter5A}. We report the flux densities (and $3\sigma$ upper limits for non-detections) and spectral indices of Ter5A after breaking up the March and April VLA images into $\sim6$min snapshots at both 2.5\,GHz and 3.5\,GHz. The midpoint of each snapshot, in MJD and binary phase, are listed in columns 1 and 2, respectively.}\\\hline\hline
MJD & phase & length & $S_{2.5\,GHz}$ & $S_{3.5\,GHz}$ & $\alpha$  \\
 (d) &      & (s)    & ($\mu$Jy)  & ($\mu$Jy) &  \\ \hline
\endfirsthead

\multicolumn{5}{l}
{ \tablename\ \thetable{} -- continued from previous page} \\\hline\hline
MJD & phase & length & $S_{2.5\,GHz}$ & $S_{3.5\,GHz}$ & $\alpha$  \\
 (d) &      & (s)    & ($\mu$Jy)  & ($\mu$Jy) &  \\
\hline
\endhead

\hline \multicolumn{5}{l}{Continued on next page}
\endfoot

\hline
\endlastfoot

59652.44022013 & 0.199 & 380 & $<101$ & $<77$ &  \\
59652.44449688 & 0.255 & 359 & $111\pm36$ & $<74$ &  \\
59652.44966262 & 0.324 & 380 & $335\pm32$ & $253\pm24$ & $-0.84\pm0.41$ \\
59652.45393967 & 0.380 & 359 & $581\pm32$ & $390\pm25$ & $-1.19\pm0.25$ \\
59652.45910541 & 0.449 & 380 & $240\pm32$ & $281\pm22$ & $+0.50\pm0.47$ \\
59652.46338246 & 0.505 & 359 & $659\pm33$ & $431\pm22$ & $-1.26\pm0.21$ \\
59652.46854819 & 0.573 & 380 & $991\pm32$ & $456\pm20$ & $-2.31\pm0.16$ \\
59652.47282495 & 0.630 & 359 & $958\pm28$ & $516\pm23$ & $-1.84\pm0.16$ \\
59652.47799070 & 0.698 & 380 & $954\pm30$ & $462\pm21$ & $-2.16\pm0.17$ \\
59652.48226774 & 0.755 & 359 & $987\pm31$ & $457\pm21$ & $-2.29\pm0.17$ \\
59652.48743347 & 0.823 & 380 & $1016\pm29$ & $459\pm21$ & $-2.36\pm0.16$ \\
59652.49171053 & 0.880 & 359 & $919\pm30$ & $423\pm21$ & $-2.31\pm0.18$ \\
59652.49687626 & 0.948 & 380 & $925\pm28$ & $477\pm20$ & $-1.97\pm0.15$ \\
59652.50115302 & 0.004 & 359 & $607\pm28$ & $336\pm21$ & $-1.76\pm0.23$ \\
59652.50631875 & 0.073 & 380 & $<86$ & $160\pm21$ & --  \\
59652.51059581 & 0.129 & 359 & $<83$ & $126\pm20$ &  -- \\
59652.51576154 & 0.198 & 380 & $<86$ & $92\pm19$ & --  \\
59652.52003859 & 0.254 & 359 & $<86$ & $130\pm21$ &  -- \\
59652.52520433 & 0.322 & 380 & $<82$ & $<56$ & --  \\
59652.52948109 & 0.379 & 359 & $<84$ & $135\pm19$ & --  \\
59652.53464682 & 0.447 & 380 & $507\pm28$ & $308\pm20$ & $-1.49\pm0.26$ \\
59652.53892387 & 0.504 & 359 & $834\pm29$ & $420\pm22$ & $-2.04\pm0.19$ \\
59652.54408961 & 0.572 & 380 & $831\pm30$ & $410\pm20$ & $-2.10\pm0.18$ \\
59652.54836665 & 0.629 & 359 & $838\pm29$ & $386\pm20$ & $-2.31\pm0.19$ \\
59652.55353239 & 0.697 & 380 & $871\pm26$ & $380\pm20$ & $-2.47\pm0.18$ \\
59652.55780915 & 0.753 & 359 & $844\pm26$ & $405\pm20$ & $-2.18\pm0.17$ \\
59652.56297487 & 0.822 & 380 & $842\pm27$ & $375\pm19$ & $-2.41\pm0.18$ \\
59652.56725193 & 0.878 & 359 & $796\pm28$ & $364\pm20$ & $-2.33\pm0.19$ \\
59652.57241766 & 0.947 & 380 & $839\pm26$ & $378\pm19$ & $-2.37\pm0.18$ \\
59652.57669471 & 0.003 & 359 & $659\pm26$ & $362\pm19$ & $-1.79\pm0.20$ \\
59652.58186044 & 0.071 & 380 & $120\pm26$ & $180\pm20$ & $+1.30\pm0.76$ \\
59652.58613720 & 0.128 & 359 & $<89$ & $130\pm20$ & --  \\
59652.59130292 & 0.196 & 380 & $<83$ & $<59$ & --  \\
59652.59557997 & 0.253 & 359 & $<84$ & $<63$ & --  \\
59652.60074570 & 0.321 & 380 & $215\pm28$ & $216\pm20$ & $+0.03\pm0.49$ \\
59652.60502275 & 0.378 & 359 & $496\pm28$ & $364\pm20$ & $-0.92\pm0.24$ \\
59652.61018849 & 0.446 & 380 & $750\pm26$ & $355\pm20$ & $-2.23\pm0.20$ \\
59652.61446524 & 0.502 & 359 & $811\pm29$ & $398\pm20$ & $-2.12\pm0.18$ \\
59652.61963097 & 0.571 & 380 & $774\pm27$ & $415\pm20$ & $-1.86\pm0.18$ \\
59652.62390801 & 0.627 & 359 & $720\pm30$ & $389\pm21$ & $-1.83\pm0.20$ \\
59652.62907375 & 0.695 & 380 & $747\pm27$ & $442\pm21$ & $-1.56\pm0.18$ \\
59652.63335080 & 0.752 & 359 & $534\pm28$ & $344\pm21$ & $-1.31\pm0.24$ \\
59652.63851652 & 0.820 & 380 & $<82$ & $<57$ & --  \\
59652.64279328 & 0.877 & 359 & $<84$ & $<61$ & --  \\
59652.64795901 & 0.945 & 380 & $<84$ & $<64$ & --  \\
59652.65223605 & 0.002 & 359 & $<89$ & $<62$ & --  \\
59652.65740179 & 0.070 & 380 & $<85$ & $<65$ & --  \\
59652.66167883 & 0.127 & 359 & $165\pm32$ & $277\pm23$ & $+1.64\pm0.66$ \\
59685.37756832 & 0.613 & 380 & $787\pm33$ & $265\pm19$ & $-3.25\pm0.25$ \\
59685.38184503 & 0.669 & 359 & $805\pm28$ & $262\pm18$ & $-3.34\pm0.23$ \\
59685.38701070 & 0.737 & 380 & $870\pm27$ & $327\pm18$ & $-2.91\pm0.19$ \\
59685.39128770 & 0.794 & 359 & $822\pm27$ & $310\pm17$ & $-2.90\pm0.19$ \\
59685.39645338 & 0.862 & 380 & $826\pm26$ & $286\pm17$ & $-3.16\pm0.20$ \\
59685.40073037 & 0.919 & 359 & $811\pm26$ & $325\pm17$ & $-2.72\pm0.18$ \\
59685.40589605 & 0.987 & 380 & $783\pm25$ & $300\pm16$ & $-2.86\pm0.19$ \\
59685.41017275 & 0.044 & 359 & $654\pm26$ & $250\pm17$ & $-2.87\pm0.24$ \\
59685.41533842 & 0.112 & 380 & $265\pm25$ & $221\pm16$ & $-0.53\pm0.36$ \\
59685.41961542 & 0.168 & 359 & $115\pm24$ & $96\pm17$ & $-0.50\pm0.88$ \\
59685.42478110 & 0.237 & 380 & $<74$ & $107\pm16$ & --  \\
59685.42905809 & 0.293 & 359 & $97\pm26$ & $82\pm17$ & $-0.40\pm1.14$ \\
59685.43422377 & 0.362 & 380 & $255\pm25$ & $143\pm16$ & $-1.73\pm0.45$ \\
59685.43850047 & 0.418 & 359 & $561\pm25$ & $234\pm17$ & $-2.61\pm0.26$ \\
59685.44366615 & 0.486 & 380 & $720\pm24$ & $243\pm16$ & $-3.24\pm0.22$ \\
59685.44794314 & 0.543 & 359 & $812\pm24$ & $281\pm17$ & $-3.16\pm0.20$ \\
59685.45310882 & 0.611 & 380 & $824\pm24$ & $266\pm16$ & $-3.37\pm0.20$ \\
59685.45738581 & 0.668 & 359 & $835\pm25$ & $278\pm17$ & $-3.28\pm0.20$ \\
59685.46255148 & 0.736 & 380 & $802\pm24$ & $269\pm16$ & $-3.25\pm0.20$ \\
59685.46682819 & 0.793 & 359 & $767\pm23$ & $239\pm15$ & $-3.48\pm0.21$ \\
59685.47199386 & 0.861 & 380 & $368\pm23$ & $166\pm15$ & $-2.38\pm0.33$ \\
59685.47627085 & 0.917 & 359 & $<74$ & $90\pm17$ & --  \\
59685.48143653 & 0.986 & 380 & $<69$ & $<50$ & --  \\
59685.48571352 & 0.042 & 359 & $98\pm23$ & $73\pm17$ & $-0.87\pm1.11$ \\
59685.49087919 & 0.110 & 380 & $<70$ & $72\pm17$ & --  \\
59685.49515590 & 0.167 & 359 & $<75$ & $<50$ & --  \\
59685.50032157 & 0.235 & 380 & $<70$ & $<50$ & --  \\
59685.50459856 & 0.292 & 359 & $<71$ & $<52$ & --  \\
59685.50976423 & 0.360 & 380 & $168\pm26$ & $177\pm17$ & $+0.20\pm0.56$ \\
59685.51404123 & 0.417 & 359 & $601\pm24$ & $247\pm17$ & $-2.65\pm0.24$ \\
59685.51920690 & 0.485 & 380 & $749\pm23$ & $307\pm17$ & $-2.66\pm0.19$ \\
59685.52348359 & 0.541 & 359 & $706\pm24$ & $267\pm17$ & $-2.90\pm0.22$ \\
59685.52864926 & 0.610 & 380 & $<73$ & $102\pm17$ & --  \\
59685.53292625 & 0.666 & 359 & $<78$ & $74\pm17$ & --  \\
59685.53809192 & 0.735 & 380 & $<76$ & $<51$ & --  \\
59685.54236892 & 0.791 & 359 & $<79$ & $<52$ &  -- \\
59685.54753459 & 0.859 & 380 & $549\pm24$ & $253\pm18$ & $-2.31\pm0.25$ \\
59685.55181129 & 0.916 & 359 & $495\pm24$ & $140\pm19$ & $-3.81\pm0.44$ \\
59685.55697696 & 0.984 & 380 & $<74$ & $<53$ & --  \\
59685.56125396 & 0.041 & 359 & $<79$ & $<55$ & --  \\
59685.56641933 & 0.109 & 380 & $<77$ & $<57$ & --  \\
59685.57069633 & 0.166 & 359 & $<81$ & $<58$ & --  \\
59685.57586227 & 0.234 & 380 & $<81$ & $<55$ & --  \\
59685.58013898 & 0.290 & 359 & $<80$ & $<56$ &  -- \\
59685.58530464 & 0.359 & 380 & $<84$ & $<56$ & --  \\
59685.58958135 & 0.415 & 359 & $<92$ & $<60$ & --  \\
59685.59474702 & 0.484 & 380 & $<90$ & $<61$ & -- \\
59685.59902401 & 0.540 & 359 & $<95$ & $<62$ & -- 
\label{tab:ter5a_flx}
\end{longtable}

%%%%%%%%%%%%%%%%%%%%%%%%%%%%%%%%%%%%%%%%%%%%%%%%%%

% Don't change these lines
\bsp	% typesetting comment
\label{lastpage}
\end{document}